# A hierarchical memory architecture overcomes context limits in long-horizon multi-agent computational modeling

Shivendra G. Tewari* and Holly Kimko

Systems Medicine, Clinical Pharmacology & Safety Sciences, AstraZeneca, Gaithersburg, Maryland

*Corresponding author: Shivendra.Tewari@AstraZeneca.com

# Abstract

Large language models (LLMs) demonstrate remarkable reasoning capabilities, yet their stateless architecture fundamentally limits deployment in long-horizon research workflows requiring multi-session continuity and quantitative rigor. Here we present *Ensemble QSP*, a multi-agent framework featuring a three-layer hierarchical memory architecture that keeps injected context bounded and constant with project duration (mid-term project state: median 301 tokens, maximum 4,050 across 104 runs) by capping each state category and evicting completed work, thereby enabling continuous autonomous operation without context degradation. The system orchestrates five specialist worker agents under domain-expert principal investigators, enforcing physical constraints through physics-based checklists and structured domain knowledge. Comprehensive benchmarking demonstrates autonomous pharmacokinetic-pharmacodynamic model selection without human intervention, improved PK parameter recovery relative to a single-agent baseline, and stable task interpretation across linguistically diverse formulations of the same request. Repeating the principal benchmarks with open-weight `DeepSeek-V4-Flash`, `DeepSeek-V4-Pro`, and `Llama 3.1 70B` configurations reproduced the principal architectural conclusions across PKPD model selection, difficult PK parameter recovery, literature retrieval and synthesis, prompt robustness, and published PBPK model implementation. Feature-level ablation during literature-based PBPK reproduction further shows that task tracking, retrieval, reasoning, memory, and PI oversight address distinct scientific and operational failure modes. These results demonstrate that the framework's principal findings are not dependent on proprietary LLMs, while the high-complexity PBPK experiments show that writer capability and run-to-run reliability remain consequential for stringent physical-consistency checks. The architecture is structurally agnostic to computational biology; adding a scientific domain requires only a new PI-agent configuration.

# Introduction

## *The context window bottleneck*

Large language models (LLMs) have advanced from text-generation tools to scientific reasoning engines capable of autonomous chemical synthesis [1], protein structure prediction [2], and end-to-end scientific discovery and publication [3]. These successes demonstrate substantial parametric knowledge and reasoning capacity for complex scientific tasks. However, a critical gap persists: real scientific projects are not single-prompt endeavors. They unfold over weeks or months, require continuity across sessions, accumulate intermediate results, and demand adherence to domain-specific physical constraints throughout.

Quantitative systems pharmacology (QSP) exemplifies this challenge [4]. A typical QSP modeling workflow involves formulating systems of stiff ordinary differential equations (ODEs), fitting parameters to heterogeneous non-clinical and clinical data from multiple studies, validating against held-out observations, and iterating across drug candidates while maintaining physiological plausibility, unit consistency, and regulatory-grade documentation. No existing LLM agent framework supports such workflows autonomously because existing frameworks were designed for task-level completion rather than project-level continuity.

Multi-agent system (MAS) frameworks such as AutoGen [5] enable role-based agent collaboration but treat context management naively: agents either pass full conversation histories, leading to $O(N \times T)$ token growth, where N is the session count and T is the number of tokens per session, or start fresh on each interaction and lose accumulated scientific knowledge. Other orchestration libraries, including CrewAI and LangChain-LangGraph, provide related role-based and graph-based workflows but are not evaluated directly here. The AI Scientist [3] demonstrated end-to-end scientific discovery and manuscript generation but operates within a single session on well-scoped machine-learning tasks, without addressing the stiff-ODE solving, multi-source data integration, and physical-constraint enforcement that characterize mathematical biology.

We address this limitation with a hierarchical memory architecture comprising three bounded layers. The short-term layer holds the live working context injected each turn: a fixed window of recent conversation turns (4–20, set by model context size), a capped working-memory `scratchpad` (≤20,000 characters for specialist agents, ≤8,000 for principal investigators) in which an agent records exact facts, file structure, parameter indices, data columns, so they survive context truncation, and a rolling buffer of the five most recent cross-agent task outcomes. The mid-term layer is a structured project-state JSON from which only bounded, task-relevant slices are injected: the five most-recent active user request[i]s (with a running total), the 20 most-recently modified files, recent jobs, milestones, and the last three session summaries. This is paired with a separate auto-summarizing decision log (capped at 16,000 characters) that records why each step was taken. The long-term layer is project-independent domain knowledge: a ~24,000-character QSP modeling handbook injected into every specialist agent, per-domain knowledge files and physics checklists, and retrieval of source-paper text and reference code during code review. Because every layer is explicitly capped and completed work is evicted, the injected mid-term state stays bounded as a project runs across sessions, estimated median of 301 tokens (interquartile range 215–478; maximum 4,050) across 104 project runs, rather than growing with the full conversation history, which was unbounded in early prototypes. This replaces the $O(N \times T)$ growth of naive history concatenation with context that is constant with respect to project duration.

# Results

## *System architecture*

The Ensemble QSP architecture addresses the context budget problem through structured state compression (Fig. 1). Instead of naively concatenating conversation history, which rapidly degrades LLM reasoning performance [6], the system maintains a categorized state (tasks, files created, issues encountered, jobs submitted, milestones achieved) with selective injection of only task-relevant fields. Decision logs are auto summarized, and domain knowledge is injected only when physics validation is triggered. This compresses a typical large-token project history into a compact-token structured state, preventing context window bloat over long-horizon projects.

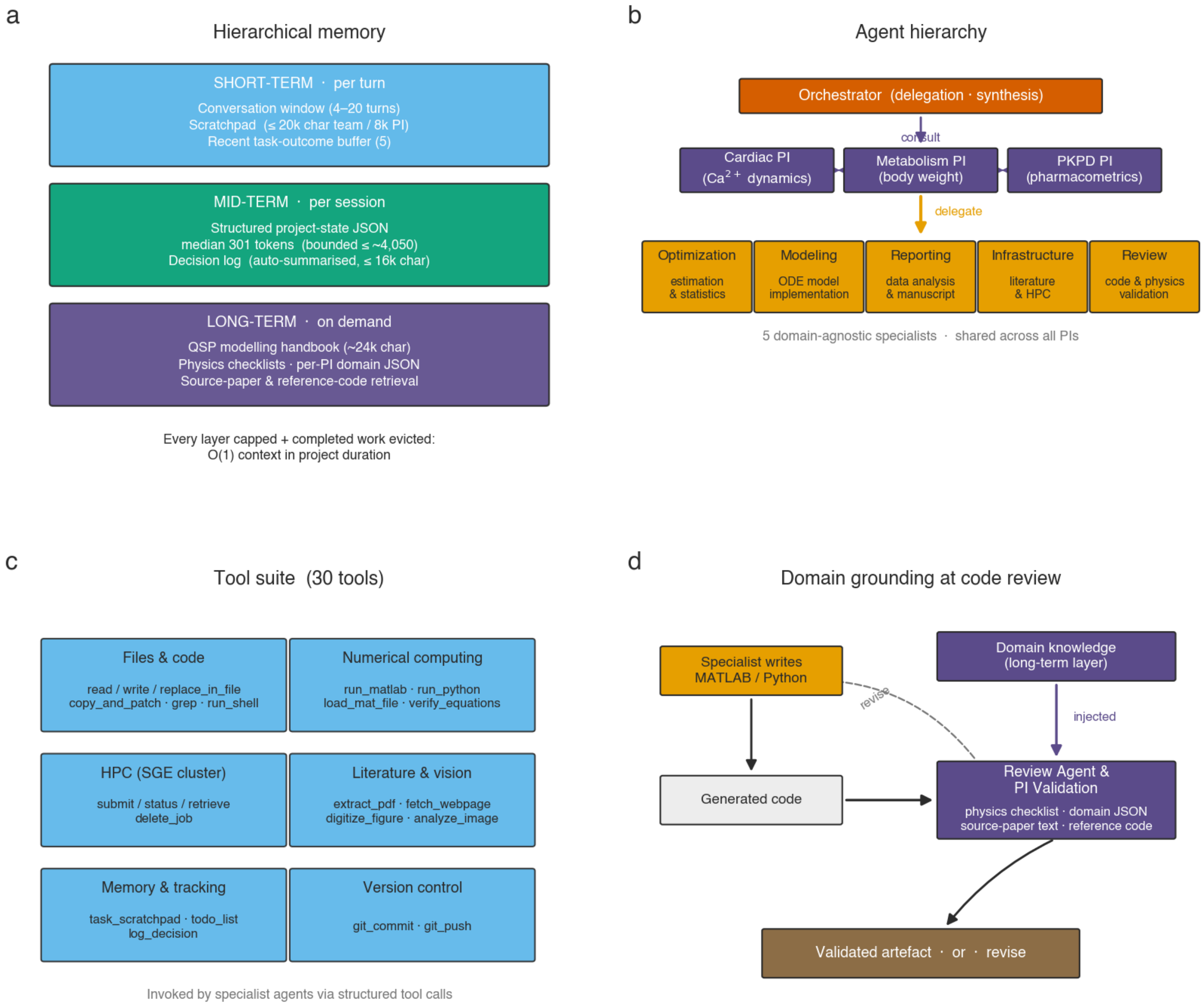

**Figure 1. System architecture. a,** Three-layer hierarchical memory comprising short-term working context (bounded conversation window and capped scratchpad, updated per turn), mid-term structured project state (JSON; median 301 tokens, updated per session), and long-term domain knowledge (handbook, physics checklists, and on-demand source-paper retrieval). **b,** Agent hierarchy in which the PI Agent delegates to five specialist sub-agents with non-overlapping roles in optimization, modeling,

reporting, infrastructure, and code review. **c,** Tool suite for literature retrieval, numerical computing, HPC, and reporting. **d,** Domain grounding through physics checklists, domain JSON files, and retrieval-augmented generation from publications during code review and validation.

A two-tier agent hierarchy organizes computation and enables iterative execution of assigned tasks [7]. Domain-expert principal-investigator (PI) agents can be configured to provide strategic oversight. In this study, three PI configurations are presented: metabolic physiology for energy balance and body-weight change, cardiac electrophysiology for calcium dynamics, and PKPD modeling for benchmarking the multi-agent and single-agent systems. Five specialist sub-agents have distinct technical responsibilities: the Optimization Agent performs parameter estimation and statistical analysis; the Modeling Agent implements mathematical models and ODE systems; the Reporting Agent analyzes data and prepares reports; the Infrastructure Agent coordinates literature retrieval and HPC execution; and the Review Agent performs code-quality and physical-consistency checks. Supplementary Note 1 provides the corresponding role map. A central orchestrator manages task delegation, tool execution, consultation, result synthesis, and parallel execution through thread pooling.

Domain grounding reduces the risk of physically implausible outputs [8, 9] through three mechanisms. First, domain-specific physics checklists—thirteen checks for cardiac calcium, fourteen for energy balance, and nine for PKPD/PBPK—are injected automatically during code review. Second, structured domain-knowledge files encode model physiology, unit conventions, parameter ranges, and documented pitfalls. Third, retrieval-augmented generation (RAG) [10] from indexed project publications allows generated equations to be compared directly with published source material during verification by the PI and Review Agents.

The architecture is structurally agnostic to computational biology: specialist agents perform optimization, ODE implementation, analysis, literature search, and code review regardless of scientific domain. Adding a new domain requires only a new PI agent configuration (system prompt, domain JSON, and physics checklist). As evidence of this modularity, the three exemplary PI agents share identical sub-agent teams and a central orchestrator; adding a PI Agent (e.g., PKPD) required less than one day of development time with zero orchestrator changes. The `BasePIAgent` class handles all LLM provider logic, while domain-specific subclasses override only the necessary routing methods.

## *Multi-agent versus single-agent benchmarking*

We evaluated the system against a single-pass LLM baseline, GitHub Copilot (GHCP), across two complementary benchmarks with known ground truth (Fig. 2). We selected GHCP as the baseline because it is widely used by computational modelers in the pharmaceutical industry.

### PKPD model selection benchmark.

Twenty synthetic datasets were generated from six candidate model structures: 2 PK × 3 PD: one-compartment and two-compartment pharmacokinetics, each paired with $E_{max}$, sigmoidal $E_{max}$, and indirect response (IDR) pharmacodynamics. The multi-agent system achieved 100% correct model selection (20/20 datasets) on the first attempt with zero human intervention, whereas GHCP achieved 70% accuracy (14/20) but required nine human-assisted iterations/loops to self-identify the best ODE solver for the PKPD task, resulting in total 9×20 (180 fitting scripts) versus 1 job (20 fits) for the MAS.

Performance differentiation was most pronounced on IDR datasets, which require solving stiff ODEs with appropriate initial conditions. The MAS correctly identified the structure, i.e., PK and PD model type, of all eight IDR datasets (100%) with a median PD parameter error of 14.1%, whereas GHCP identified only five (62%) with a median error of 68.4% on those correctly selected. For the simpler $E_{max}$ models, both systems performed comparably (MAS: 8.3% median error; GHCP: 5.0%), however, note that GHCP's lower median error reflects 100-fold tighter solver tolerances perhaps adopted during its nine debugging iterations so not an equivalent comparison. See Supplementary Figures 1 and 2 for detailed comparison.

## PK parameter recovery benchmark.

Twenty difficult two-compartment subcutaneous PK profiles identified from 100 synthetic PK datasets were fitted (see Supplementary Figure 3 and Methods). The MAS recovered parameters within 10% of ground truth for 70% of profiles (14/20), compared with 20% for GHCP. Total wall-clock time was longer for the MAS (105 minutes versus 18–25 minutes), reflecting the overhead of multi-agent coordination. As a result of longer computing time, the number of successful ground-truth parameter value recoveries per unit time for GHCP are higher than MAS, but total number of ground truth recoveries are higher for MAS. See Supplementary Figure 4 for a head-to-head comparison.

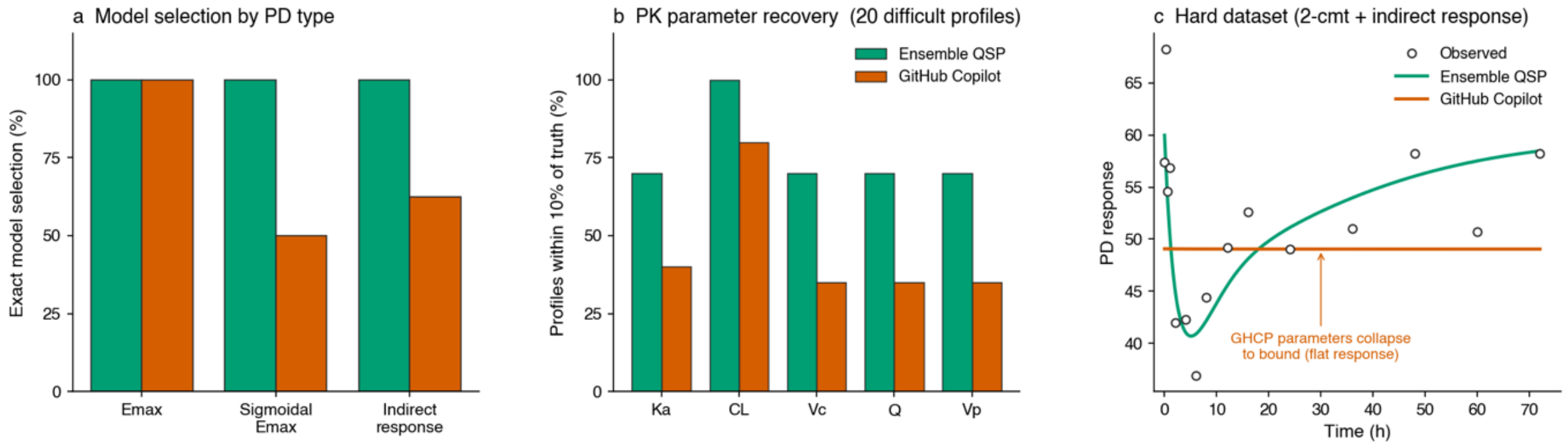

**Figure 2. Multi-agent versus single-agent benchmarking. a,** PKPD model-selection accuracy across 20 synthetic datasets, stratified by PD model type: direct $E_{max}$, sigmoidal $E_{max}$, and indirect response (IDR). The MAS achieved 100% accuracy on the first attempt, whereas GHCP required nine self-debugging iterations. **b,** Fraction of difficult PK profiles for which all parameters were recovered within 10% of ground truth by the MAS and GHCP. **c,** Representative fits for a difficult IDR dataset showing the MAS simulation (green), GHCP simulation (orange), and data (open circles). GHCP estimates of the IDR parameters $k_{in}$, $k_{out}$, and $I_{max}$ remained near their lower bounds, reducing both the drug-driven perturbation and the relaxation rate; consequently, the predicted PD response remained near baseline and produced the flat trajectory.

## Open-weight models reproduce the principal benchmark conclusions.

A central question for reproducible scientific deployment is whether the observed performance depends on proprietary frontier models or arises from the architectural scaffolding. We therefore repeated the most challenging subsets of the controlled PKPD model-selection and PK parameter-recovery benchmarks using open-weight models while holding the multi-agent architecture, domain knowledge, code-review workflow, and PI-directed oversight constant. The evaluated configurations used `DeepSeek-V4-Flash` or `DeepSeek-V4-Pro` as the writer and `Llama 3.1 70B` for code review; Table 1 and the archived run metadata report the complete role assignments. This design directly tests independence from proprietary models and sensitivity to capability differences among open-weight writers. The subsequent body-weight and cardiac-electrophysiology case studies retain their original closed-weight configurations because they illustrate domain-specific PI workflows rather than provide evidence for the cross-model reproducibility benchmark.

For the PKPD model-selection benchmark, we re-evaluated the eleven most difficult datasets, comprising all IDR and two-compartment sigmoidal-$E_{max}$ cases that most strongly stress structural model identification. All three configurations, `DeepSeek-V4-Flash`, `DeepSeek-V4-Pro`, and the frontier `Claude Opus 4.8` writer, correctly identified both the PK and PD structures for all eleven datasets (11/11 exact model selection). `DeepSeek-V4-Flash` closely matched `Claude Opus 4.8` in parameter recovery: median absolute parameter error was 6.8% versus 7.0%, mean error was 15.7% versus 16.2%, and maximum error was 92.4% versus 90.9%, respectively, across 77 parameter comparisons. In contrast, `DeepSeek-V4-Pro` had a higher median error of 9.0% and a markedly poorer error distribution. An extreme $E_{max}$ estimate in one dataset increased its mean error to 2,465.0% and maximum error to 140,522.3%, despite retaining exact structural model selection. Thus, the lower-cost Flash configuration reproduced the frontier writer's structural and parameter-recovery performance, whereas the Pro configuration provided no accuracy advantage and was substantially more vulnerable to extreme parameter outliers.

We next evaluated the open-weight configuration on the twenty most difficult pharmacokinetic profiles from the parameter-recovery benchmark. These profiles were specifically selected because they represent severe identifiability challenges arising from flip-flop kinetics, $K_a$-$V_c$ tradeoffs, Q-$V_p$ non-identifiability, and convergence to local minima. The open-weight `DeepSeek-V4-Pro` system recovered all five parameters within 10% of ground truth for 14 of 20 profiles (70%), matching the performance previously observed with the frontier-model configuration. Across all 100 fitted parameter values, 82% were recovered within 10% of their true value. Individual parameter recovery rates were 70% for $K_a$, 100% for CL, 70% for $V_c$, 85% for Q, and 85% for $V_p$.

Together, these results demonstrate that the principal benchmark conclusions are not restricted to a proprietary LLM family. When coupled with structured project memory, domain grounding, code review, and PI-directed corrective feedback, open-weight configurations reproduced the exact structural model selection and difficult PK parameter recovery observed with the frontier configuration. However, the extreme parameter outlier produced by `DeepSeek-V4-Pro` and the variable success of `DeepSeek-V4-Flash` in diagnosing and correcting the high-complexity Shah–Betts mass-balance issue, with successful reproduction in two of three independent runs, show that writer capability and run-to-run reliability remain consequential for particular quantitative and physical-consistency tasks. The evidence therefore supports architectural robustness across model families without implying that the underlying LLM is interchangeable for every task.

***Table 1. Token usage and cost for open-weight and frontier code-writer configurations***

| Agent (role) | Model | Input tokens | Output tokens | Calls | Cost (US$) |
|---|---|---|---|---|---|
| **Open-weight Flash writer: 11/11 exact** | | | | | |
| Optimization Agent (Writer) | `DeepSeek-V4-Flash` | 6,796,536 | 44,159 | 93 | 1.31 |
| Review Agent (Review) | `Llama 3.1 70B` | 47,077 | 562 | 1 | 0.03 |
| Infrastructure Agent (HPC) | `DeepSeek-V4-Pro` | 450,107 | 4,087 | 8 | 0.80 |
| PI Agent (Oversight) | `DeepSeek-V4-Pro` | 902,115 | 44,547 | 29 | 1.72 |
| **Subtotal** | | **8,195,835** | **93,355** | **131** | **3.86** |
| **Open-weight Pro writer: 11/11 exact** | | | | | |
| Optimization Agent (Writer) | `DeepSeek-V4-Pro` | 5,876,560 | 51,532 | 67 | 10.40 |
| Review Agent (Review) | `Llama 3.1 70B` | 52,336 | 8,192 | 1 | 0.04 |
| PI Agent (Oversight) | `DeepSeek-V4-Pro` | 332,306 | 16,359 | 10 | 0.64 |
| **Subtotal** | | **6,261,202** | **76,083** | **78** | **11.08** |
| **Frontier writer: 11/11 exact** | | | | | |
| Optimization Agent (Writer) | `Claude Opus 4.8` | 2,866,761 | 21,805 | 25 | 16.36 |
| Reporting Agent (Analysis) | `DeepSeek-V4-Pro` | 300,271 | 5,370 | 5 | 0.54 |
| Review Agent (Review) | `Llama 3.1 70B` | 95,564 | 8,914 | 2 | 0.08 |
| PI Agent (Oversight) | `Gemini 3.1 Pro` | 273,614 | 1,666 | 10 | 0.57 |
| **Subtotal** | | **3,536,210** | **37,755** | **42** | **17.55** |

The cost comparison reinforces this distinction (Table 1). The `DeepSeek-V4-Flash` configuration completed the benchmark for $3.86 ($0.35 per dataset), approximately 78% less than the `Claude Opus 4.8` configuration at $17.55 ($1.60 per dataset), while producing nearly identical median, mean, and maximum parameter errors (see Supplementary Figure 5 inset). By contrast, the `DeepSeek-V4-Pro` configuration cost $11.08 ($1.01 per dataset) yet exhibited a substantially worse parameter-error distribution, driven by an extreme outlier (Dataset 14), and therefore offered a less favorable cost-performance tradeoff than either Flash or Opus. These results identify the Flash configuration—not the more expensive Pro configuration—as the strongest open-weight alternative for this benchmark.

These results demonstrate that the framework is not restricted to proprietary frontier models, but they also show that open-weight model choice matters. When paired with structured project memory, domain grounding, code review, and PI-directed auto-looping, `DeepSeek-V4-Flash` achieved frontier-comparable structural selection and parameter recovery at substantially lower cost. `DeepSeek-V4-Pro` also retained exact structural selection, but its poorer parameter-error distribution and higher cost relative to Flash indicate that a nominally more capable writer model does not necessarily provide a better scientific or economic outcome within the same architecture.

## *Architecture-component ablation during literature-based PBPK model reproduction*

To determine which components of the architecture support reliable scientific implementation with open-weight language models, we performed feature-level ablation during autonomous reproduction of the atenolol minimal physiologically based pharmacokinetic (mPBPK) model reported by Kir et al. [11]. This four-state ODE model provides a controlled literature-to-code benchmark that nevertheless requires faithful retrieval of published parameters, correct implementation of multi-phase oral absorption, adherence to a forward-simulation-only task specification, and iterative code review. All ablation runs used the same source publication, digitized reference profiles, simulation request, and open-weight agent configuration; only the indicated architectural component was disabled (Fig. 3b–c). Supplementary Figure 6 places this low-complexity task in a broader complexity context by showing an independent `DeepSeek-V4-Flash` reproduction of Figure 4 from Kir et al. alongside `DeepSeek-V4-Pro` and `DeepSeek-V4-Flash` reproductions of the substantially larger 93-state ODE model reported by Shah and Betts [12].

**Full-system reproduction.** With all architecture components enabled, the open-weight system reproduced the published atenolol concentration-time profiles for the intravenous reference condition and the control and malnourished oral conditions (Fig. 3a). The ablation analysis assessed not only terminal goodness of fit but also compliance with the requested scientific workflow, quantitative fidelity to the publication, and operational efficiency. The full system achieved a mean $R^2$ of 0.604 across the three scenarios and completed the task in 39.7 min (Fig. 3b).

**Structured task tracking prevents scope drift.** When structured task tracking was disabled (`no_todo_list`), the agent deviated from the explicit forward-simulation request and authored an unrequested parameter-fitting script (`fit_atn_absorption.m`). This failure did not arise from an inability to write executable code; rather, it reflected loss of the task constraint as the workflow progressed, demonstrating that persistent task state is required to preserve scientific scope across iterative implementation and review. Notably, this ablation achieved the highest mean $R^2$ (0.768) because parameter fitting improved agreement with the digitized data. However, it constitutes a workflow failure rather than a scientific improvement because the system no longer performed the requested forward-simulation task (Fig. 3b-c).

**Publication retrieval anchors quantitative fidelity.** Ablating retrieval of the source publication (`no_rag`) led to parameter miscalculations and overprediction of the oral peak concentration: $C_{max}$ was 30.92 µg/mL compared with the 25.0 µg/mL reference value, corresponding to a +23.7% error. The implementation retained moderate aggregate agreement with the digitized data (mean $R^2$ = 0.647), but it did not reproduce the published parameterization correctly. Thus, general domain knowledge and agent reasoning were insufficient to reconstruct paper-specific quantitative details reliably without direct access to the publication (Fig. 3b-c).

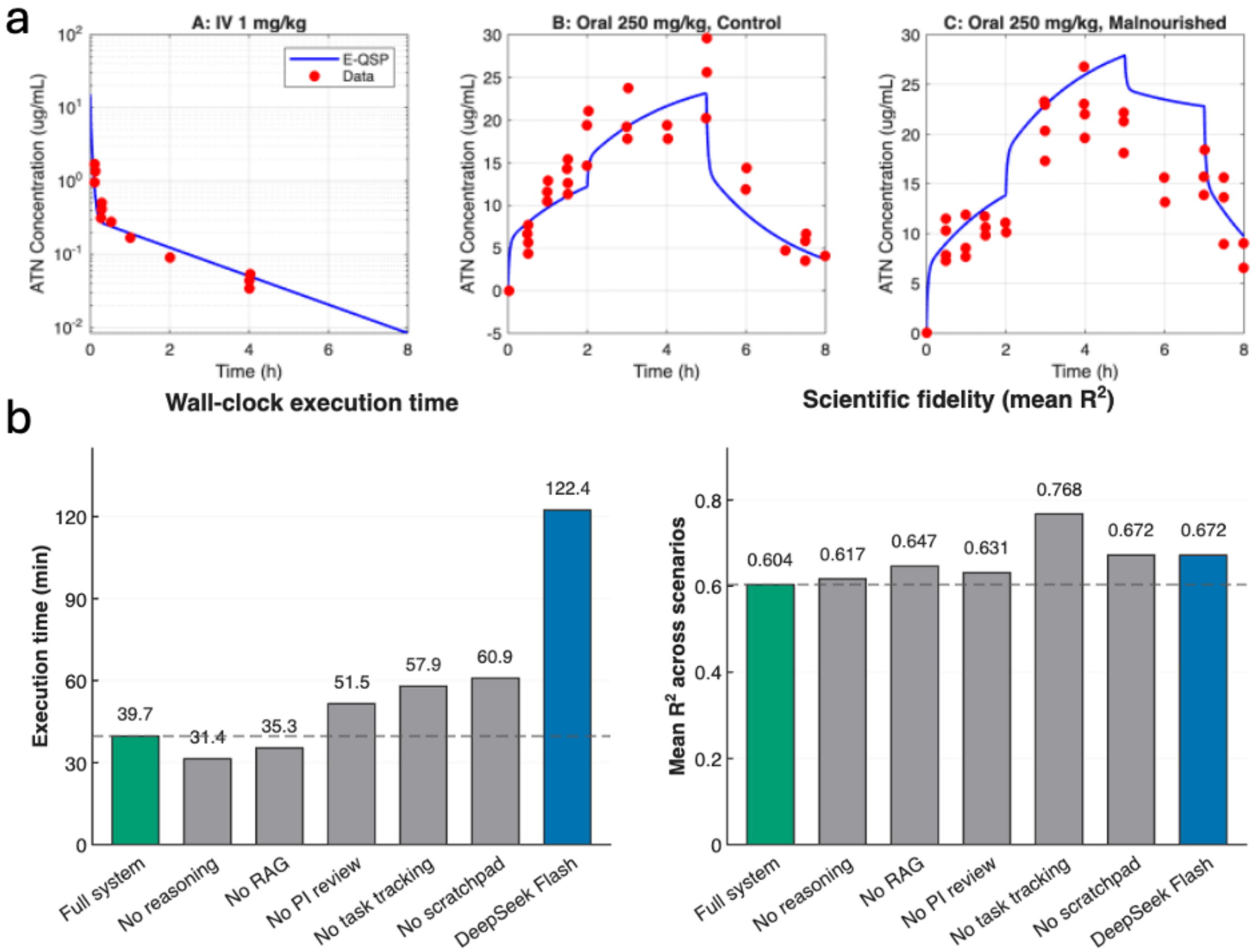

c **Failure modes and workflow effects**

| Run / Variant | Fidelity preserved? | Operational penalty? | Failure mode / Workflow effect |
|---|---|---|---|
| **Full system** | Yes | No | Reference (DeepSeek Pro) |
| No reasoning | No | No | Piecewise absorption logic error |
| No RAG | No | No | Parameter extraction error |
| No PI review | No | No | Reduced quantitative fidelity |
| No task tracking | No | Yes | Simulation turned parameter fitting |
| No scratchpad | Yes | Yes | Longer execution, accuracy preserved |
| **DeepSeek Flash** | Yes | Yes | Accuracy preserved, more review cycles |

**Figure 3. Architecture-component ablation during autonomous literature-based model reproduction. a,** Autonomous reproduction of the atenolol mPBPK model of Kir et al. [11]. Ensemble QSP simulations (lines) are compared with manually digitized Figure 4 data (closed circles) for intravenous 1 mg/kg, oral 250 mg/kg control, and oral 250 mg/kg malnourished scenarios. **b,** Scientific fidelity (mean $R^2$ across the three reproduced scenarios) and wall-clock execution time across architecture-component ablations. **c,** Failure modes and workflow effects associated with each ablation, illustrating the distinct scientific and operational functions provided by memory, retrieval, reasoning, task tracking, PI oversight, and writer-model capability.

**Extended reasoning supports piecewise mathematical implementation.** Disabling extended reasoning (`no_thinking_mode`) produced errors in the timing logic for the multi-phase absorption ODE. Although the resulting code executed, it reproduced the malnourished scenario poorly ($R^2$ = 0.3850), below the predefined passing threshold, and yielded a mean $R^2$ of 0.617 across scenarios. This ablation distinguishes syntactic executability from scientific correctness: explicit reasoning was required to reconcile phase-specific absorption intervals with the published equations and parameter table (Fig. 3b-c).

**PI oversight improves quantitative fidelity.** Removing the PI review layer (`no_pi_agent`; labeled "No PI oversight" in Fig. 3) did not remove implementation capability, but it eliminated the scientific review and corrective-feedback step. The resulting workflow completed in 51.5 min and achieved a mean $R^2$ of 0.631, with reduced quantitative fidelity relative to the intended full review process. This result indicates that PI oversight contributes primarily through scientific validation, consistency checking, and corrective guidance rather than direct code generation (Fig. 3b–c).

**Agent scratchpad memory primarily affects operational efficiency.** Removing the agent scratchpad memory layer (`no_agent_memory`) preserved overall scientific fidelity (mean $R^2$ = 0.672) but increased execution time to 60.9 min, compared with 39.7 min for the full system. The implementation required additional review and corrective iterations, indicating that the scratchpad supports operational efficiency and continuity of reasoning rather than determining terminal numerical accuracy (Fig. 3b–c).

**Open-weight writer capability primarily determines operational efficiency.** Replacing the primary code writer with the lower-cost `DeepSeek-V4-Flash` model preserved scientific fidelity (mean $R^2$ = 0.672) and successfully reproduced all three experimental scenarios, but increased wall-clock execution time approximately threefold (122.4 versus 39.7 min) and required additional review and correction cycles. The architecture therefore compensated for reduced writer-model capability through iterative review and PI-directed feedback, trading operational efficiency for a preserved final scientific outcome (Fig. 3b-c).

Together, these ablations reveal graded degradation patterns across scientific fidelity, operational efficiency, and workflow correctness. Structured task tracking preserves scientific intent, publication retrieval anchors paper-specific quantitative details, extended reasoning supports translation of mathematical descriptions into executable code, PI oversight provides scientific review and corrective feedback, and hierarchical memory improves operational continuity. No single component explains system performance; reliable autonomous model implementation instead emerges from the coordinated operation of the architectural layers.

## *Linguistic invariance and prompt robustness*

Autonomous deployment requires system behavior to depend on the scientific content of a request rather than on its surface phrasing. The architecture addresses this requirement by enriching each request with relevant project-memory context and normalizing it through the PI Agent. We evaluated two properties of this normalization layer—linguistic invariance and robustness to prompt specificity—using Dataset 11, a benchmark task with a known two-compartment PK plus direct $E_{max}$ PD ground truth (Supplementary Notes 6 and 7).

**Linguistic invariance.** We issued the same fully specified modeling task through nine linguistically diverse rephrasing while holding the scientific request constant. The PI Agent correctly parsed and delegated all nine variants on the first attempt, but the stochastic optimization outcomes were not uniformly identical. In the initial combined run, seven variants selected the ground-truth two-compartment PK plus $E_{max}$ PD model. The terse-imperative variant selected a one-compartment PK structure, and the stream-of-consciousness variant selected two-compartment PK plus sigmoidal $E_{max}$. The passive-academic variant selected the correct structure but produced an extreme mean parameter error of 63,151%, driven by implausibly scaled PK estimates. Repeating the identical terse-imperative and passive-academic prompts recovered the ground-truth structure with mean/median parameter errors of 20.1%/16.5% and 13.4%/10.3%, respectively. The stream-of-consciousness result remained a model-discrimination difference rather than a parsing failure: the sigmoidal $E_{max}$ model had AIC 42.91 versus 45.42 for the direct $E_{max}$ model ($\Delta$AIC = 2.51), exceeding the Burnham–Anderson threshold of 2 but only modestly favoring the more complex model. Its estimated sigmoidicity coefficient was close to unity (approximately 1.3), making the fitted sigmoidal form mechanistically close to direct $E_{max}$ ($\gamma = 1$). Thus, eight of nine variants recovered the ground-truth structure (including the two reruns), whereas one selected a closely related nested PD model. These results support robust task interpretation across writing styles while showing that stochastic optimization, parameter scaling, and discrimination between nearly nested PD models remain consequential (Supplementary Figures 7–15).

**Prompt robustness.** Does specificity of an ambiguous task matter? A complementary test varied the completeness of the task specification across eight prompts, ranging from a 1,383-character expert brief to a 134-character bare instruction. Independent rescoring of the archived outputs corrected two false negatives from the original ablation parser: both the novice-vague and minimal prompts selected the ground-truth two-compartment PK plus $E_{max}$ PD model. The expert-verbose prompt initially selected an incorrect PK structure but recovered the ground-truth model when the identical prompt was rerun. The final analysis therefore yielded exact model selection for seven of eight variants; the remaining Scandinavian-formal run selected a two-compartment sigmoidal $E_{max}$ model whose AIC differed from the ground-truth $E_{max}$ model by only 1.7, below the conventional threshold for meaningful model discrimination. Across the seven exact-match results, mean parameter error averaged 16.0% (median 15.0%; range 5.4%–26.7%). Notably, the novice-vague and minimal prompts achieved mean parameter errors of 5.4% and 25.1%, respectively, showing that prompt brevity or incomplete terminology did not preclude correct model identification. Together, the two tests show that the architecture is robust to how a well-posed task is phrased and can infer the intended model-comparison workflow from sparse requests, while numerical optimization and model discrimination retain expected stochastic variability (Supplementary Figures 16–23).

## *Literature pipeline benchmarking*

A critical prerequisite for autonomous modeling is the ability to locate, extract, and synthesize information accurately from scientific literature without human curation. We evaluated the end-to-end literature pipeline across retrieval, extraction, and synthesis using 17 test cases spanning cardiac electrophysiology, metabolic regulation, and PKPD (Table 2).

**Retrieval.** Ten identical queries (R01–R10) targeted known published papers across diverse modeling domains (population PK [13–15], PBPK [11, 12], indirect-response PD [16], tumor growth inhibition [17], bone-remodeling QSP [18, 19], and clinical trial reports [20–26]). The open-weight configuration retrieved the intended literature for all ten queries on the first attempt (10/10), whereas the closed-weight configuration succeeded initially for seven queries and required author-anchored refinements for R02, R06, and R09 (7/10 first pass; 10/10 after refinement). For the nine open-weight runs unaffected by infrastructure failure, mean wall-clock time was 1,004 s (16.7 min; range 217–2,984 s). The recorded 18,558-s duration for the Shah–Betts query (R05) was excluded from this representative timing summary because the target paper was identified and downloaded within the first minute; subsequent completion was delayed by a transient Azure API 403 error, tool timeouts, and repeated continuation of text extraction and summarization rather than by retrieval complexity. The observed R05 elapsed time is retained in Supplementary Table 3 for transparency.

**Extraction with hallucination resistance.** Five cases (E01–E05) tested structured model-detail extraction from paper text, incorporating adversarial cases designed to elicit hallucination. The system correctly identified irrelevant or mismatched records, including the off-domain articles used as adversarial controls [27, 28], rather than fabricating modeling content and accurately extracted model details from the genuine Shah and Betts PBPK paper. All five cases were adjudicated as correct by the PI Agent (5/5 behavioral pass). Mean wall-clock time was 226 s (3.8 min; range 96–459 s), based on the recorded run durations; see Supplementary Table 4 for details.

**Multi-paper synthesis.** Two synthesis scenarios tested the most demanding tier of the literature pipeline: autonomous retrieval of multiple related publications, reconciliation of differences in model structure and parameterization, and production of comparative scientific artifacts. In both scenarios, the open-weight configuration produced a more complete and scientifically informative synthesis than the earlier closed-weight configuration. In S01, the system retrieved three semaglutide population-PK analyses spanning subcutaneous, intravenous, and oral administration, rather than limiting the comparison to the two target papers. It correctly distinguished absolute from apparent disposition parameters, reconciled the one- and two-compartment subcutaneous models, and recognized that the oral model fixed systemic disposition to the subcutaneous and intravenous estimates. The synthesis identified subcutaneous bioavailability of approximately 85% versus 0.8% for oral semaglutide, an approximately 106-fold difference, and correctly attributed the major formulation contrast to absorption and bioavailability rather than to formulation-dependent systemic clearance. It also extracted the principal covariate findings, including the effects of body weight, post-dose fasting time, and water volume on oral exposure. These findings were supported by a structured comparison of model architecture, study populations, parameter estimates, covariates, and clinical implications across the retrieved studies.

In S02, the open-weight system expanded the requested comparison of at least three FcRn-enabled monoclonal-antibody PBPK models to ten publications spanning foundational whole-body models, minimal and nested-endosome formulations, platform models, engineered-antibody extensions, and models incorporating plasma-phase FcRn binding [12, 29, 30]. The resulting synthesis traced the progression from equilibrium FcRn binding and transcytosis to explicit pH-dependent endosomal trafficking, target binding, half-life extension, competitive FcRn inhibition, and plasma-phase FcRn interactions. It compared tissue and sub-compartment structures, estimated ODE dimensionality, predicted human IgG half-life, cross-species scalability, and the principal innovation of each model. Importantly, it maintained calibrated confidence by labeling ODE counts as estimates when they were not stated explicitly and by distinguishing full-text evidence from abstract-only sources. The synthesis identified a cross-model consensus of approximately 20–23 days for typical human IgG1 half-life and showed how model complexity increased from minimal hypothesis-testing systems to 15-tissue platform models with detailed endosomal trafficking.

Both open-weight syntheses were manually verified and outperformed the prior closed-weight runs primarily in breadth, cross-paper reconciliation, and evidentiary transparency: S01 added a third relevant population-PK analysis and a formulation-independent disposition interpretation, whereas S02 expanded a three-paper comparison to a ten-model historical and mechanistic synthesis. These results indicate that the literature pipeline is not dependent on proprietary frontier models and that, for retrieval-intensive synthesis, the open-weight configuration can exploit the architecture's structured search, project memory, and PI-directed verification to produce more comprehensive outputs than the closed-weight configuration. Across the two synthesis tasks, mean wall-clock time was 543 s for the open-weight runs versus 942 s for the closed-weight runs—a reduction of 399 s (42%) and an approximately 1.7-fold increase in speed (Supplementary Table 4).

**Table 2: Literature pipeline benchmark results**

| Tier | Cases (n) | Result | Traps | Mean time (s) |
|---|---|---|---|---|
| Retrieval (first pass) | 10 | Open: 10/10 first pass; closed: 7/10 first pass, 10/10 after refinement | N.A. | 1,004* |
| Extraction (model-detail) | 5 | 5/5 correct | 3/5 (all passed) | 226 |
| Synthesis (multi-paper) | 2 | 2/2 Manually verified | N.A. | 543 (open); 942 (closed) |

## *Autonomous QSP parameter fitting*

The following application case studies illustrate the domain-specific capabilities of the Body-Weight and Cardiac-Electrophysiology PI Agents. They were generated using the original closed-weight configuration and were not included in the procedural open-weight re-evaluation, which focused on the controlled benchmarks described above.

### Body-weight management case

We applied the full system to autonomous body-weight modeling for GLP-1 receptor agonist (GLP-1RA) drugs using the Hall 2024 body-composition QSP framework [31]. To simulate GLP-1RA effects, we added two receptor-occupancy functions: appetite suppression, which reduces caloric intake, and homeostatic-feedback attenuation, which modulates the counter-regulatory response to weight loss. When prompted, the system retrieved trial publications through optimized PubMed queries, extracted body-weight time-course data from published figures using vision-language models, configured a PK model with trial-specific dose-escalation schedules, estimated pharmacodynamic parameters using the Hall model, and validated the fits against observed clinical endpoints for semaglutide, liraglutide, and dulaglutide in sequence.

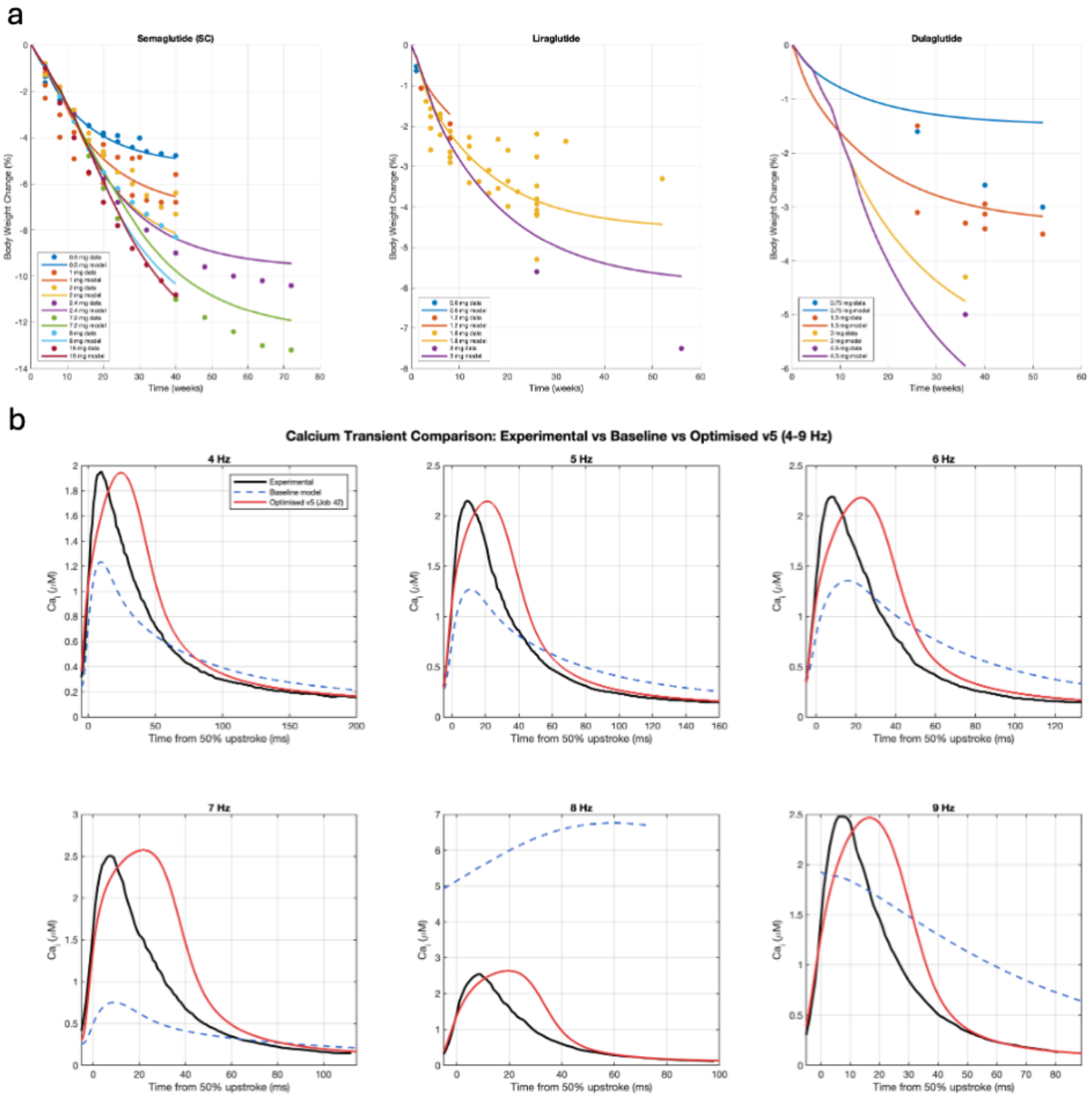

**Figure 4: Autonomous model identification and parameter fitting. a** Hall model [31] coupled with drug-specific PK models. Model-predicted (lines) versus observed (symbols) percentage body weight change for semaglutide (blue, 11 cohorts, 0.5–16 mg), liraglutide (orange, 8 cohorts, 0.6–3.0 mg), and dulaglutide (green, 11 cohorts, 0.75–4.5 mg) from the combined 5-parameter fit. Each cohort was initialized by their representative body weight, height, and sex distribution. **b** Gattoni et al. [32] cardiac electrophysiology model. The system iteratively formulated and tested structural modifications to the baseline model to explain the frequency-dependent acceleration of relaxation (FDAR) observed in isolated rat ventricular myocytes. Model-predicted intracellular calcium transients for the baseline (dashed lines) and the optimized CaMKII-coupled variant (red lines) are compared against experimental recordings (black lines) across pacing frequencies from 4 to 9 Hz. The identified model successfully captures both the positive amplitude staircase and the accelerated decay kinetics at higher pacing rates.

**Empirical dose-specific semaglutide data fitting.** In a first autonomous analysis phase, the system independently estimated four PD parameters: $E_{max,GLP,app}$ (maximum appetite suppression), $EC_{GLP,app}$ (concentration at 50% of maximum appetite, $EC_{50}$), $E_{max,k_{prop}}$ (maximum feedback suppression), and $EC_{k_{prop}}$ (feedback $EC_{50}$), at each of the seven semaglutide dose levels (0.5–16 mg) across five trial programs (SUSTAIN-1, SUSTAIN-7, SUSTAIN-FORTE, STEP-UP T2D, and PIONEER PLUS; 11 cohorts total). Since the system was not provided a PK model, the system autonomously retrieved semaglutide prescribing information from the FDA database and implemented a standard analytical one-compartment PK representation that is easily extendable to other compounds with available FDA labels. This approach directly utilized the FDA-reported terminal half-life and steady-state exposure proportionality, ensuring fidelity to established clinical exposures without the need for *de novo* structural model fitting. Based on the user's request, the system also validated their one-compartment PK implementation with the published two-compartment model of Overgaard et al. [15], see Supplementary Note.

As expected, dose-specific fits achieved an RMSE < 0.85% body weight change across all 11 cohorts, with the lowest residuals at high doses (RMSE = 0.10–0.11% at 7.2–16 mg) and the highest at 1.0 mg (RMSE = 0.63–0.84%), reflecting inter-trial heterogeneity at this dose level. The system identified two key parameter trends: $E_{max,GLP,app}$ increased from 0.28 at 0.5 mg to 0.49 at 16 mg, suggesting a dose-dependent efficacy ceiling empirically; and $EC_{GLP,app}$ converged to the lower bound (∼100 pmol/L free) at low doses, indicating near-maximal receptor occupancy. $EC_{k_{prop}}$ varied over two orders of magnitude (2,500–200,000 pmol/L), which the system correctly flagged as poorly identifiable. During a subsequent domain-expert review, this lack of identifiability is likely driven by the relatively short duration of many included trials, whereas the homeostatic feedback effect ($k_{prop}$) primarily manifests in the extended time course required to reach a weight-loss plateau.

**Autonomous parameter estimation.** Based on dose-specific insights, the system autonomously designed and executed a combined 5-parameter fit across three structurally distinct GLP-1RAs (Fig. 4a): semaglutide (11 cohorts, 0.5–16 mg), liraglutide (8 cohorts, 0.6–3.0 mg), and dulaglutide (11 cohorts, 0.75–4.5 mg). The parameterization comprised drug-specific appetite $EC_{50}$ values and shared GLP1-efficacy parameters ($E_{max,app}$, $EC_{k_{prop}}$), with $E_{max,k_{prop}}$ fixed at 0.5 based on the sensitivity analysis and literature [31]. Parameter estimation used `particleswarm` and converged to the following $EC_{GLP,app}$ values: 222 pmol/L (semaglutide), 616 pmol/L (liraglutide), 3888 pmol/L (dulaglutide), and shared $E_{max,app}$ and $EC_{k_{prop}}$ values of 0.384 and 1000 pmol/L (at lower bound), respectively. The resulting $EC_{50}$ ranking on a free-concentration basis (semaglutide < liraglutide < dulaglutide) correctly recapitulates known clinical potency ordering.

## Cardiac electrophysiology case

**Autonomous model mechanism extension.** To test whether the system could autonomously extend a QSP model, we applied the MAS to a structurally distinct problem: frequency-dependent acceleration of relaxation (FDAR) in rat ventricular myocytes. The system received the 19-state, 64-parameter rat ventricular-myocyte model of Gattoni et al. [32], validated at 6 Hz, together with experimental calcium-transient features—amplitude, time to peak, and 50% and 90% relaxation times ($TD_{50}$ and $TD_{90}$)—across pacing frequencies of 4–9 Hz.

The experimental data [33] show a positive $Ca^{2+}$ amplitude staircase, with peak $Ca^{2+}$ increasing from 1.81 to 2.43 μM, and pronounced FDAR, with $TD_{90}$ shortening from 82.5 ms at 4 Hz to 43.1 ms at 9 Hz. The baseline model reproduced neither behavior and became numerically unstable above its validated frequency of 6 Hz: it predicted non-physiological relaxation prolongation ($TD_{90}$ > 240 ms at 9 Hz) and failed to relax at frequencies above 6 Hz (Fig. 4b).

Rather than treating the 64 parameters as fixed, the system serially hypothesized, implemented, and tested structural additions to the model, drawing on the cardiac electrophysiology literature to identify candidate frequency-sensing mechanisms. The system converged on a CaMKII-dependent regulatory module: active CaMKII—a frequency-dependent integrator of the calcium signal—accelerates SR calcium re-uptake by up-regulating SERCA activity through a Hill-type activation function, complemented by refinements to ryanodine-receptor release. This extended the model from 64 to 70 parameters and introduced biophysical coupling absent from the baseline formulation, providing a candidate mechanistic basis for FDAR in which faster pacing elevates CaMKII activity, which in turn speeds relaxation.

Fitting the extended model with `particleswarm` reproduced both the amplitude staircase and FDAR across all six frequencies, tracking peak $Ca^{2+}$ increase from 1.80 to 2.52 μM (simulated) and the near-halving of $TD_{90}$ from 79.7 to 34.8 ms (simulated); fit quality was highest at lower frequencies and degraded modestly at the upper end of the range (0.13–0.15 at 8–9 Hz). Notably, parameters were estimated independently at each frequency rather than as a single shared set; the resulting frequency-dependent parameter shifts, concentrated in the CaMKII-SERCA pathway, provide direct evidence for the proposed relaxation mechanism. A unified parameterization, in which one parameter set adapts to pacing rate through the CaMKII feedback loop itself, is an ongoing extension of this work.

# Discussion

The results demonstrate that hierarchical memory management is a central enabler of long-horizon autonomous scientific computation with LLMs. Controlled ablation across a complexity gradient further shows that memory, task tracking, retrieval, reasoning, PI oversight, and writer-model capability address distinct scientific and operational failure modes; no single component or prompting strategy is sufficient.

**Architecture components address distinct failure modes.** The open-weight atenolol ablation study shows that reliable literature-based model reproduction is not explained by a single prompting or model-capability effect. Structured task tracking preserved the requested scope and prevented an unsolicited shift from forward simulation to parameter fitting. Publication retrieval anchored paper-specific parameter values and prevented a 23.7% $C_{max}$ error. Extended reasoning was required to implement multi-phase absorption timing correctly, whereas the lower-tier writer ultimately reached comparable numerical accuracy only through substantially more time and review cycles. These findings support a division of labor within the architecture: memory and task state preserve intent, retrieval supplies source-specific evidence, reasoning supports mathematical translation, and review loops compensate for lower-capability writers at an operational cost.

**Architectural versus model-capability contributions.** The linguistic-invariance and prompt-robustness experiments isolate the role of the PI Agent's normalization layer from downstream numerical variability. For linguistic invariance, all nine paraphrases were parsed and delegated correctly, but only seven selected the ground-truth structure in the initial combined run. Identical reruns corrected the terse-imperative structural miss and the passive-academic extreme parameter-scaling error; the stream-of-consciousness run instead favored the nested sigmoidal $E_{max}$ model by $\Delta AIC = 2.51$, with $\gamma$ close to 1. Prompt robustness yielded exact model selection for seven of eight specification variants after independent rescoring and one identical rerun; the remaining Scandinavian-formal run selected sigmoidal $E_{max}$ with $\Delta AIC = 1.7$, indicating no meaningful separation from direct $E_{max}$. Together, these findings show that architectural normalization reduces sensitivity to surface phrasing and prompt completeness, but does not eliminate stochastic optimization failures or ambiguity between closely related model structures.

**Scientific reasoning in the GLP-1RA case study.** The clinical-fitting workflow extended beyond parameter optimization: the system retrieved supporting FDA label data, implemented an explicit one-compartment linearized PK model that could be applied across compounds, and performed sensitivity analyses to identify parameters governing appetite suppression and homeostatic feedback. Fixing the maximal GLP-1RA effect on homeostatic feedback reduced confounding between drug-specific appetite suppression and attenuation of homeostatic feedback during weight loss.

This multi-step reasoning chain—spanning literature retrieval, hypothesis generation, model revision, and validation—was executed with minimal human intervention across multiple sessions, enabled by the hierarchical memory architecture's preservation of scientific rationale across context boundaries. Initiated in January 2026, the project has continued for several months while testing multiple mechanistic hypotheses. Its compressed project state remains approximately 4,000 tokens, demonstrating sustained operational continuity without context-window exhaustion. This long-running case complements the controlled benchmarks by showing that the bounded project-state representation can preserve decisions and intermediate results over an extended research workflow.

**Open-weight reproducibility across task classes.** The procedural re-evaluation extended beyond a single sensitivity analysis: open-weight configurations reproduced the difficult PKPD model-selection subset, the hardest PK parameter-recovery profiles, linguistic-invariance and prompt-robustness tests, literature retrieval and multi-paper synthesis, and published PBPK models spanning four-state and 93-state systems. Performance was not uniformly determined by nominal model tier. `DeepSeek-V4-Flash` matched the frontier writer on the difficult PKPD subset and completed the low-complexity atenolol implementation. In triplicate attempts on the higher-complexity Shah–Betts model [12], `DeepSeek-V4-Flash` diagnosed and corrected the mass-balance issue sufficiently to reproduce Figure 8 in two of three runs, whereas `DeepSeek-V4-Pro` produced the successful reproduction shown in Supplementary Figure 6b. These results support the central architectural claim while defining a capability boundary: open-weight models are sufficient for the principal findings, but reliable diagnosis and correction of physical-consistency issues becomes more challenging as model dimensionality and inter-compartmental coupling increase.

**Comparison with existing frameworks.** AutoGen [5], CrewAI, and LangChain-LangGraph provide multi-agent orchestration but lack persistent project state and domain-specific grounding. The AI Scientist [3] demonstrates research automation within single sessions on machine-learning tasks. ChemCrow [34] integrates chemistry tools with an LLM in a single-agent system but does not provide long-horizon, cross-session continuity. To our knowledge, Ensemble QSP is the first framework to combine autonomous scientific implementation, cross-session hierarchical memory, and domain-expert grounding for multi-month projects validated on published mathematical models. The literature-pipeline benchmark further differentiates the architecture: in all three adversarial cases, the system correctly declined to fabricate content from irrelevant or unretrievable publications. This behavior reflects PI-Agent oversight at the extraction stage and provides an explicit scientific-integrity safeguard absent from single-agent retrieval systems without domain-expert adjudication.

**Limitations.** Although the principal benchmarks were reproduced with open-weight `DeepSeek-V4-Flash`, `DeepSeek-V4-Pro`, and `Llama 3.1 70B` configurations, reproducibility still depends on preserving exact model checkpoints, inference settings, prompts, software dependencies, random seeds, and benchmark artifacts. Open-weight access reduces dependence on changing commercial APIs but does not eliminate sensitivity to checkpoint revisions, deployment infrastructure, or stochastic optimization. Reported wall-clock task-processing times should not be interpreted as direct measures of intrinsic LLM performance because they also reflect network bandwidth, API availability and rate limits, transient service outages, tool timeouts, and shared compute or HPC queue latency. We therefore log model identifiers and API parameters for every session and deposit the benchmark inputs, outputs, prompts, and run metadata with the code. The domain-grounding layer also requires expert curation of physics checklists, which is a one-time investment per domain. Validation is currently confined to pharmacology, although the addition of three PI-agent domains without orchestrator changes provides structural evidence of extensibility. Finally, the GitHub Copilot comparison represents one competing approach; broader comparisons with AutoGen and CrewAI on identical tasks were not feasible because of access and configuration constraints.

**Implications.** The demonstrated ability to autonomously reproduce published models and fit clinical data across 30 cohorts and three drug classes with minimal human intervention suggests a near-term role for memory-augmented agent systems as computational research assistants. In the evaluated workflows, the system completed routine model implementation, fitting, validation, and reporting tasks within hours while maintaining scientific review through automated checks and PI-directed feedback. Because the architecture operates with open-weight models, organizations could deploy the language models on local or institution-managed infrastructure and thereby eliminate per-call API charges, although hardware acquisition, energy use, maintenance, and inference operations would remain. Lastly, the architecture is intended to augment rather than replace domain experts by reducing the time required for reproducible routine analyses and allowing experts to focus on model assumptions, interpretation, and decision-making.

# Methods

**Language models and gateway access.** Model assignments varied by benchmark and are therefore reported at the run level rather than consolidated under a single configuration. Models were accessed through an institutional AI Gateway that routed requests to AWS Bedrock, Azure OpenAI, or Google endpoints. Gateway access was used for operational convenience and does not make the architecture dependent on a proprietary model service: the open-weight models can be downloaded and deployed independently, and the prompts, run metadata, code, and model assignments required to reproduce the reported workflows are archived in Zenodo. The original MAS-versus-GHCP PKPD and PK comparisons used `Claude 4.6 Opus` as the shared writer model (`us.anthropic.claude-opus-4-6-v1`; AWS Bedrock), as described in Supplementary Notes 2 and 3. The later 11-dataset writer-model comparison used `Claude Opus 4.8` as the frontier writer (`us.anthropic.claude-opus-4-8`; AWS Bedrock), `Gemini 3.1 Pro Preview` for PI oversight (`gemini-3.1-pro-preview`; Google native route), `DeepSeek-V4-Pro` for reporting (`DeepSeek-V4-Pro`; Azure OpenAI), and `Llama 3.1 70B Instruct` for code review (`us.meta.llama3-1-70b-instruct-v1:0`; AWS Bedrock), as reported in Table 1. The procedural revision additionally evaluated open-weight configurations using `DeepSeek-V4-Flash` and `DeepSeek-V4-Pro` for writer, PI, reporting, and infrastructure roles as specified for each benchmark, with `Llama 3.1 70B Instruct` for code review. Reproducibility runs used the pinned `deepseek-v4-flash-0731` endpoint (Azure OpenAI); no pinned `DeepSeek-V4-Pro` variant was available, so the gateway's rolling alias was used and this limitation is recorded in the archived run metadata. The stable `Claude Sonnet 4.5` endpoint was `us.anthropic.claude-sonnet-4-5-20250929-v1:0` (AWS Bedrock). Provider route, exact model identifier, role assignment, prompts, inference parameters, timestamps, and token usage are logged for each archived run. Pricing per $10^6$ input/output tokens was \$0.19/\$0.51 for `DeepSeek-V4-Flash`, \$1.74/\$3.48 for `DeepSeek-V4-Pro`, \$0.72/\$0.72 for `Llama 3.1 70B Instruct`, \$2.00/\$12.00 for `Gemini 3.1 Pro Preview`, and \$5.50/\$27.50 for `Claude Opus 4.8`.

**Agent orchestration.** The multi-agent system is implemented in Python 3.11+ with `concurrent.futures.ThreadPoolExecutor` for parallel agent execution. Agent communication follows a structured message protocol with typed fields. The PI Agent's delegation logic routes tasks based on type, complexity, and required domain knowledge. Three autonomy levels (Interactive, Semi-Autonomous, Autonomous) govern escalation conditions. All benchmark tests were performed using the 'Autonomous' mode.

**Ablation study design.** Architectural features were independently toggled through Boolean flags in the `AblationConfig` data class. For literature-based model reproduction, ablations were evaluated using the atenolol mPBPK model of Kir et al. [11] under a fixed source publication, digitized reference dataset, simulation request, and open-weight agent configuration. The tested conditions removed structured task tracking (`no_todo_list`), publication retrieval (`no_rag`), extended reasoning (`no_thinking_mode`), PI review and corrective feedback (`no_pi_agent`), or the agent scratchpad memory layer (`no_agent_memory`). A separate configuration replaced the primary code writer with `DeepSeek-V4-Flash` while retaining the remaining architecture. Outcomes included adherence to the requested forward-simulation task, $R^2$ against manually digitized profiles, $C_{max}$ error, wall-clock time, and code-review iterations. The PKPD writer-model comparison separately used the 11 most challenging datasets and assessed exact PK and PD structure selection and parameter recovery.

**PKPD model-selection benchmark.** Twenty datasets with known ground truth were generated from six model structures (one- and two-compartment intravenous-bolus PK, each paired with direct Emax, sigmoidal Emax, or indirect-response Type IV PD). All profiles used a 100 mg IV bolus dose sampled at 15 time points spanning 0–72 h (0, 0.25, 0.5, 1, 2, 4, 6, 8, 12, 16, 24, 36, 48, 60, 72 h). Ground-truth parameters were fixed per dataset (not randomly sampled) across the ranges: CL: 0.5–4.5 L/h; $V_c$: 5–25 L; Q: 0.3–2.5 L/h (two-compartment structures only); $V_p$: 8–30 L (two-compartment only); $E_{max}$: 70–200; $EC_{50}$: 1.0–8.0; gamma: 1.5–3.5 (sigmoidal Emax only); $k_{in}$: 6–20; $k_{out}$: 0.05–0.4 $h^{-1}$; $I_{max}$: 2–10; $IC_{50}$: 1.0–8.0 (indirect-response only). Synthetic PK and PD observations each included 10% proportional residual error (multiplicative Gaussian noise on the noise-free simulated trajectory).

**PK parameter-recovery benchmark**. A separate pool of 100,000 synthetic two-compartment subcutaneous PK profiles was generated to probe parameter identifiability independently of model-selection difficulty. Each profile simulated a single 1 mg SC dose with dense hourly sampling over a 336 h (14-day) horizon; the five governing parameters ($K_a$, CL, $V_c$, Q, $V_p$) were drawn independently per profile by log-uniform sampling over ranges representative of a large-molecule SC biologic ($K_a$: 0.005–0.200 $h^{-1}$; CL: 0.005–0.250 L/h; $V_c$: 1.0–15.0 L; Q: 0.01–3.0 L/h; $V_p$: 1.0–20.0 L). From an initial multi-start fitting pass on the first 100 of the 100,000 profiles (see Supplementary Figure 3), each profile was graded by its worst single-parameter percentage error (across $K_a$, CL, $V_c$, Q, $V_p$; A: <1%, B: <5%, C: <25%, D: <100%, F: ≥100%) and classified by failure mode: $K_a$–$V_c$ swap, Q–$V_p$ non-identifiability, or convergence to a spurious local minimum. The 20 profiles graded D or F, the most severe identifiability failures, were retained as the hard-profile subset reported in the PK parameter-recovery results (Fig. 2b).

**Linguistic invariance and prompt robustness design.** Two complementary prompt sets targeted the same benchmark PKPD task with known ground truth (Dataset 11). For *linguistic invariance*, the identical, fully specified task was rephrased into nine styles while scientific content was held constant. For *prompt robustness*, eight variants altered task completeness and specificity. Outputs were scored independently with `score_pkpd_benchmark.py`. Exact match required the ground-truth two-compartment PK plus direct $E_{max}$ PD structure. Because sigmoidal $E_{max}$ nests direct $E_{max}$ at $\gamma = 1$, alternative sigmoidal selections were also interpreted using ΔAIC and the estimated $\gamma$. The terse-imperative and passive-academic linguistic-invariance prompts and the expert-verbose prompt-robustness variant were rerun unchanged to assess run-to-run optimization variability. Initial and rerun outcomes are reported separately rather than replacing the original results. All linguistic-invariance and prompt-robustness runs used `DeepSeek-V4-Flash` as the writer model.

**Literature retrieval benchmark.** Ten identical queries spanning cardiac electrophysiology, metabolic regulation, and PKPD were tested against known target papers using both open- and closed-weight configurations. First-pass success was defined as retrieval of the intended literature without modifying the original query. The open-weight configuration succeeded for all ten queries on the first attempt. The closed-weight configuration initially missed R02, R06, and R09; these searches were then systematically refined with specific author names, after which the intended literature was retrieved. Wall-clock summaries for the open-weight runs exclude R05 from the representative mean because its recorded duration was dominated by a transient API 403 error and repeated continuation after tool timeouts; the full observed elapsed time remains reported separately for transparency. For this run, the Infrastructure Agent was `DeepSeek-V4-Pro`.

**GLP-1RA clinical data sources and digitization.** Semaglutide, liraglutide, and dulaglutide body-weight-time data were either digitized from published figures or obtained from the CODEx® database. Automated figure digitization used the `Claude Sonnet 4.5` vision-language model (`us.anthropic.claude-sonnet-4-5-20250929-v1:0`), with structured validation against reported summary statistics by the Review Agent. `Qwen3-VL` was available to the PI Agent for visual verification of fitting outputs and other task artifacts when requested; it was not used by the optimized `digitize_figure` method and was not evaluated as part of the procedural open-weight benchmark. Pharmacokinetic parameters were sourced from FDA prescribing information and published population-PK analyses.

**Body weight QSP model.** The Hall body-composition model [31] comprises a 6-state ODE system (adaptive thermogenesis, glycogen, fat mass, lean mass, extracellular water, linear trend). GLP-1RA pharmacodynamic effects enter via two occupancy functions operating on free drug concentration: (1) appetite suppression: $Intake(t)=appetite_factor \times Intake_{init} - k_{prop,eff} \times (BW(t)-BW_{init})$, where the appetite factor is modulated by receptor occupancy $occ_{app}=c_{free}/(EC_{GLP,app}+c_{free})$; and (2) homeostatic feedback attenuation via an analogous $k_{prop}$ occupancy function. Dose-escalation PK was modeled as piecewise exponential approach to steady state with trial-specific titration schedules. These application results were generated using the original closed-weight configuration and are presented to illustrate the domain-specific PI workflow, not as part of the open-weight reproducibility benchmark.

**Ventricular myocyte calcium model.** The baseline model was the 19-state, 64-parameter rat ventricular-myocyte ODE system of Gattoni et al. [32], which couples action-potential generation with calcium cycling across the cytosol and sarcoplasmic reticulum, including dynamic troponin buffering, and was validated at 1 and 6 Hz. Experimental calcium-transient features—peak amplitude, time to peak, $TD_{50}$, and $TD_{90}$—were extracted at each of six pacing frequencies from 4 to 9 Hz and fitted with `particleswarm`; parameters were estimated independently at each frequency rather than as a single shared set. A unified parameterization in which one parameter set adapts to pacing rate through the CaMKII feedback loop is an ongoing extension. These application results were generated using the original closed-weight configuration and are presented to illustrate the domain-specific PI workflow, not as part of the open-weight reproducibility benchmark.

**Simulation environment.** Parameter-estimation jobs were submitted to a Sun Grid Engine (SGE)-managed cluster through automated job submission and retrieval by the Infrastructure Agent. The ablation study used fixed data seeds with varying LLM seeds across conditions. All ODE systems were solved in MATLAB® R2025b. ODE solvers, tolerances, and optimization functions were selected by the system according to the modeling task, as described in `modeling_handbook.py`. Exact package versions and software dependencies are provided in the Zenodo archive.

**Code availability.** The complete Ensemble QSP platform, including user and agent prompts, domain-knowledge files, physics checklists, benchmark datasets, the ablation harness, prompt-robustness variants, and scripts required to reproduce the open-weight evaluations, is publicly available in Zenodo at https://doi.org/10.5281/zenodo.22816553. Each archived run includes the model identifier and role assignment, project-state record, prompts, inference metadata, generated code, logs, and output artifacts required to audit the reported results.

**Data availability.** Preclinical and clinical data used in this study were derived from the cited publications or obtained from the Certara CODEx database. The synthetic benchmark datasets and ground truth, digitized published data, fitted parameters, open-weight and frontier-model benchmark outputs, project-state JSON files, token and timing logs, prompt variants, literature-retrieval and synthesis outputs, PBPK reproduction artifacts, and data underlying all main and supplementary figures and tables are publicly available at Zenodo. Access conditions for Certara

CODEx data are governed by the database license and will be described in the repository documentation.

# Acknowledgments

We gratefully acknowledge Dr. Kevin D. Hall for his conceptual guidance regarding the disentanglement of appetite suppression and homeostatic feedback mechanisms in the GLP-1RA QSP analysis. We also gratefully acknowledge Prof. James Zou for his valuable feedback on the manuscript.

# Competing Interests

SGT and HK are employees of AstraZeneca and hold shares/share options in AstraZeneca.

# Declaration of generative AI and AI-assisted technologies in the writing process

During the preparation of this work, the authors used ChatGPT for formatting checks, tense consistency, grammar checks, and language polishing. After using this tool or service, the authors reviewed and edited the content as needed and take full responsibility for the content of the publication.

# References

1. Boiko, D.A., et al., *Autonomous chemical research with large language models.* Nature, 2023. **624**(7992): p. 570-578.
2. Lin, Z., et al., *Evolutionary-scale prediction of atomic-level protein structure with a language model.* Science, 2023. **379**(6637): p. 1123-1130.
3. Lu, C., et al., *The ai scientist: Towards fully automated open-ended scientific discovery.* arXiv preprint arXiv:2408.06292, 2024.
4. Azer, K. and J.S. Barrett, *Quantitative system pharmacology as a legitimate approach to examine extrapolation strategies used to support pediatric drug development.* CPT Pharmacometrics Syst Pharmacol, 2022. **11**(7): p. 797-804.
5. Wu, Q., et al., *Autogen: Enabling next-gen llm applications via multi-agent conversation.* arXiv preprint arXiv:2308.08155, 2023.
6. Liu, N.F., et al., *Lost in the middle: How language models use long contexts.* Transactions of the association for computational linguistics, 2024. **12**: p. 157-173.
7. Hong, S., et al. *MetaGPT: Meta programming for a multi-agent collaborative framework.* in *International Conference on Learning Representations*. 2024.
8. Poweleit, E.A., A.A. Vinks, and T. Mizuno, *Artificial Intelligence and Machine Learning Approaches to Facilitate Therapeutic Drug Management and Model-Informed Precision Dosing.* Ther Drug Monit, 2023. **45**(2): p. 143-150.
9. McKenna, N., et al. *Sources of hallucination by large language models on inference tasks*. in *Findings of the association for computational linguistics: EMNLP 2023*. 2023.

10. Lewis, P., et al., *Retrieval-augmented generation for knowledge-intensive nlp tasks.* Advances in neural information processing systems, 2020. **33**: p. 9459-9474.
11. Kir, F., S. Sahin, and W.J. Jusko, *Minimal Physiologically-Based Pharmacokinetic Modeling of Atenolol and Metoprolol Absorption in Malnourished Rats.* Eur J Drug Metab Pharmacokinet, 2025. **50**(3): p. 251-263.
12. Shah, D.K. and A.M. Betts, *Towards a platform PBPK model to characterize the plasma and tissue disposition of monoclonal antibodies in preclinical species and human.* J Pharmacokinet Pharmacodyn, 2012. **39**(1): p. 67-86.
13. Yang, X.D. and Y.Y. Yang, *Clinical Pharmacokinetics of Semaglutide: A Systematic Review.* Drug Des Devel Ther, 2024. **18**: p. 2555-2570.
14. Overgaard, R.V., et al., *Clinical Pharmacokinetics of Oral Semaglutide: Analyses of Data from Clinical Pharmacology Trials.* Clin Pharmacokinet, 2021. **60**(10): p. 1335-1348.
15. Overgaard, R.V., et al., *Population Pharmacokinetics of Semaglutide for Type 2 Diabetes.* Diabetes Ther, 2019. **10**(2): p. 649-662.
16. Dayneka, N.L., V. Garg, and W.J. Jusko, *Comparison of four basic models of indirect pharmacodynamic responses.* J Pharmacokinet Biopharm, 1993. **21**(4): p. 457-78.
17. Simeoni, M., et al., *Predictive pharmacokinetic-pharmacodynamic modeling of tumor growth kinetics in xenograft models after administration of anticancer agents.* Cancer Res, 2004. **64**(3): p. 1094-101.
18. Peterson, M.C. and M.M. Riggs, *Predicting nonlinear changes in bone mineral density over time using a multiscale systems pharmacology model.* CPT Pharmacometrics Syst Pharmacol, 2012. **1**(11): p. e14.
19. Peterson, M.C. and M.M. Riggs, *A physiologically based mathematical model of integrated calcium homeostasis and bone remodeling.* Bone, 2010. **46**(1): p. 49-63.
20. Overgaard, R.V., et al., *Liraglutide 3.0 mg for Weight Management: A Population Pharmacokinetic Analysis.* Clin Pharmacokinet, 2016. **55**(11): p. 1413-1422.
21. Wilding, J.P., et al., *Exposure-response analyses of liraglutide 3.0 mg for weight management.* Diabetes Obes Metab, 2016. **18**(5): p. 491-9.
22. Rubino, D.M., et al., *Effect of Weekly Subcutaneous Semaglutide vs Daily Liraglutide on Body Weight in Adults With Overweight or Obesity Without Diabetes: The STEP 8 Randomized Clinical Trial.* JAMA, 2022. **327**(2): p. 138-150.
23. Garvey, W.T., et al., *Two-year effects of semaglutide in adults with overweight or obesity: the STEP 5 trial.* Nat Med, 2022. **28**(10): p. 2083-2091.
24. Feng, P., et al., *A Phase 1 Multiple Dose Study of Tirzepatide in Chinese Patients with Type 2 Diabetes.* Adv Ther, 2023. **40**(8): p. 3434-3445.
25. Aronne, L.J., et al., *Continued Treatment With Tirzepatide for Maintenance of Weight Reduction in Adults With Obesity: The SURMOUNT-4 Randomized Clinical Trial.* JAMA, 2024. **331**(1): p. 38-48.
26. Nicze, M., et al., *Resistant and Refractory Obesity: The Complexity of Anti-Obesity Therapy Failure.* Int J Mol Sci, 2026. **27**(6).
27. Thompson, T., et al., *1,25-dihydroxyvitamin D3 enhances the apoptotic activity of MDM2 antagonist nutlin-3a in acute myeloid leukemia cells expressing wild-type p53.* Mol Cancer Ther, 2010. **9**(5): p. 1158-68.
28. Ye, W., et al., *Cloning and functional analysis of autophagy-related gene 7 in Bombyx mori, silkworm.* Arch Insect Biochem Physiol, 2021. **107**(4): p. e21827.

29. Li, L., et al., *Simulation of monoclonal antibody pharmacokinetics in humans using a minimal physiologically based model.* AAPS J, 2014. **16**(5): p. 1097-109.
30. Hu, S., A. Datta-Mannan, and D.Z. D'Argenio, *Physiologically Based Modeling to Predict Monoclonal Antibody Pharmacokinetics in Humans from in vitro Physiochemical Properties.* MAbs, 2022. **14**(1): p. 2056944.
31. Hall, K.D., *Physiology of the weight-loss plateau in response to diet restriction, GLP-1 receptor agonism, and bariatric surgery.* Obesity (Silver Spring), 2024. **32**(6): p. 1163-1168.
32. Gattoni, S., et al., *The calcium-frequency response in the rat ventricular myocyte: an experimental and modelling study.* J Physiol, 2016. **594**(15): p. 4193-224.
33. Janssen, P.M., L.B. Stull, and E. Marbán, *Myofilament properties comprise the rate-limiting step for cardiac relaxation at body temperature in the rat.* Am J Physiol Heart Circ Physiol, 2002. **282**(2): p. H499-507.
34. Bran, A. M., et al., *Augmenting large language models with chemistry tools.* Nat Mach Intell, 2024. **6**(5): p. 525-535.

# Supplementary Information for "A hierarchical memory architecture overcomes context limits in long-horizon multi-agent computational modeling"

By

Shivendra G. Tewari* and Holly Kimko

*Corresponding author: Shivendra.Tewari@astrazeneca.com

# Table of Contents

# Supplementary Note 1. Nomenclature Mapping: Personas to Roles

To facilitate naturalistic collaboration and reinforce strict role-playing constraints within the large language models (LLMs), the agents in the Ensemble QSP framework were internally initialized with anthropomorphic personas. However, for scientific clarity and consistency in the main manuscript, these personas are referred to by their formal functional roles. Readers reviewing the raw system prompts in the sections below or in the public Zenodo archive will observe the use of these internal system names, which map to the manuscript's functional roles as follows:

- **Prof. Sneyd / Prof. Hall / Prof. Kimko:** Principal Investigator (PI) Agent provides scientific supervision, task delegation, and synthesis. The framework initiates one domain-specific PI per project: Prof. Sneyd (cardiac / calcium-dynamics), Prof. Hall (obesity / body-weight regulation), and Prof. Kimko (PKPD / PBPK and clinical-trial simulation).
- **Maya:** Optimization Agent (Parameter estimation, sensitivity analysis, and statistical evaluation).
- **Alex:** Modeling Agent (Mathematical modeling and ordinary differential equation (ODE) implementation).
- **Sam:** Infrastructure Agent (High-performance computing (HPC) coordination, tool execution, and literature retrieval).
- **Jake:** Review Agent (Code quality assurance, syntax checking, and physical constraint verification).
- **Sarah:** Reporting Agent (Data analysis, visualization, and regulatory documentation).

# Supplementary Note 2. PKPD Benchmarking Results

We compared parameter recovery across twenty synthetic PKPD datasets for two configurations that share the same frontier language model (`Claude 4.6 Opus`) but differ in architecture: the full multi-agent system (MAS) and a single-assistant baseline using GitHub Copilot (GHCP). Supplementary Figs. 1 and 2 show estimated (y-axis) versus ground-truth (x-axis) parameter values for the MAS and GHCP, respectively; circles denote parameters recovered under a correctly selected structural model and 'X' markers denote an incorrectly selected model. The MAS selected the correct PK and PD model for all 20 datasets, with estimates clustered along the identity line. In contrast, the GHCP mis-selected the structural model for 6/20 datasets, and these errors propagated into large parameter biases, most severely for the IDR parameters (e.g. $I_{max}$, IC50, $EC_{50}$). Because both configurations used the same LLM, the improvement is attributable to the multi-agent architecture rather than to the underlying model's capability.

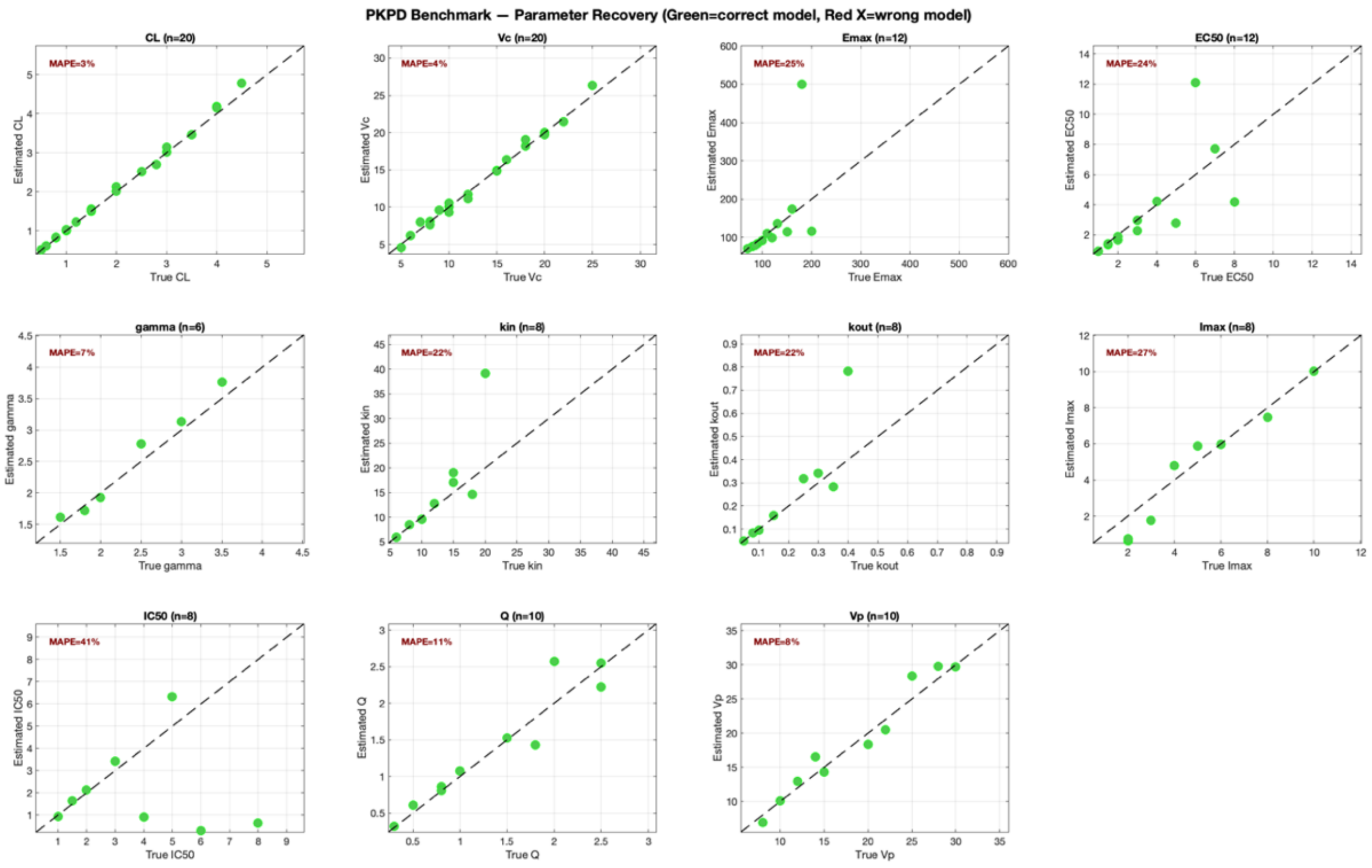

**Supplementary Figure 1:** MAS parameter recovery from twenty synthetic PKPD datasets.

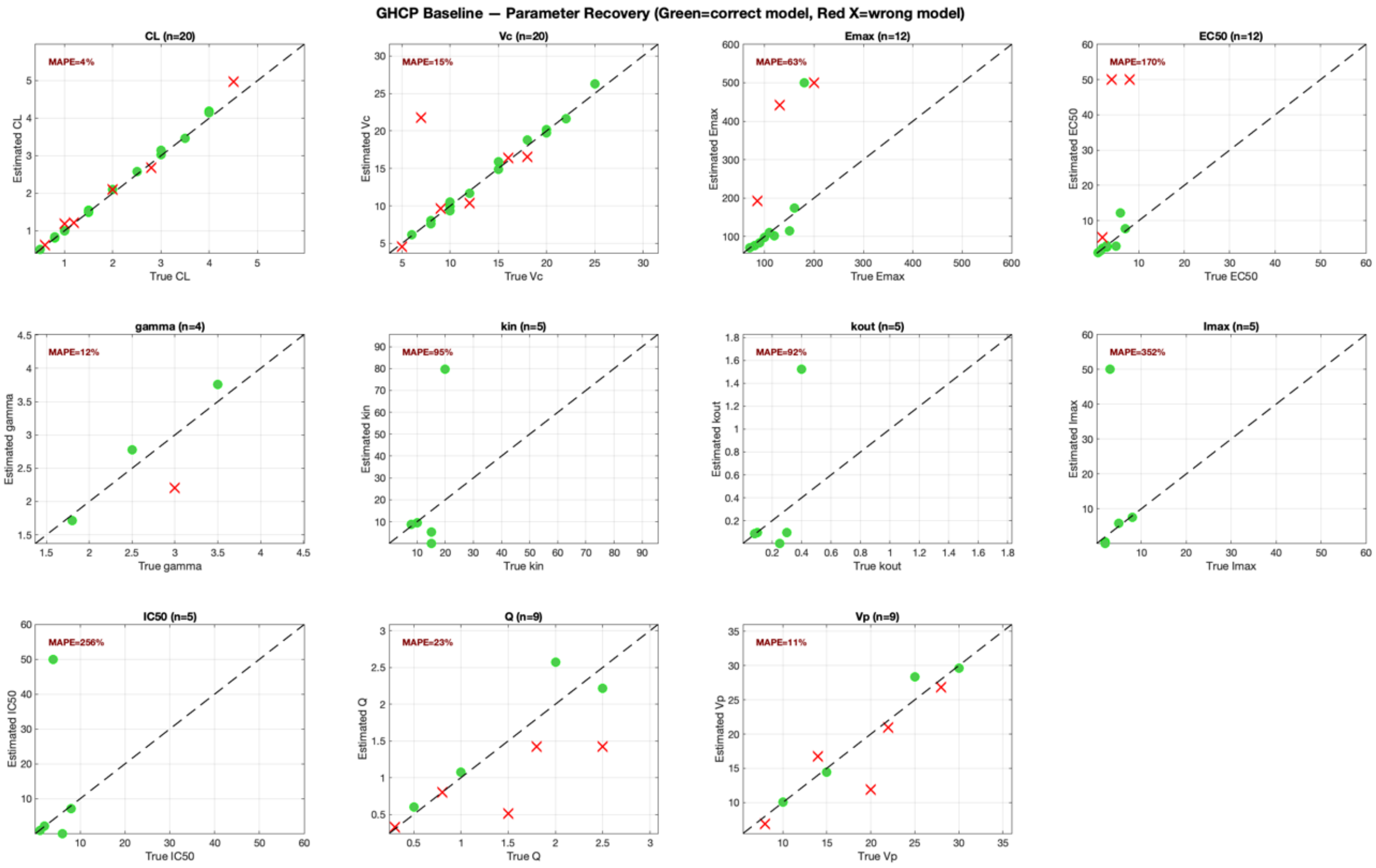

**Supplementary Figure 2:** GHCP baseline parameter recovery from twenty synthetic PKPD datasets.

# Supplementary Note 3. PK Benchmarking Results

## *Long-horizon autonomous PK fitting*

We first evaluated the system's ability to perform sustained autonomous parameter estimation using a pool of 100,000 synthetic two-compartment pharmacokinetic profiles with known ground-truth parameters ($K_a$, CL, $V_c$, Q, and $V_p$; see Methods). The MAS was instructed to write each MATLAB estimation script from scratch and began processing the pool in a continuous autonomous session. The benchmark was terminated after successful completion of the first 100 profiles to limit API cost; these 100 profiles required approximately 6 h and 129 task delegations, with a median of 2.6 min per profile. All 100 scripts executed without crashes and passed an eight-point automated code-quality lint. Recovery was excellent: the median error was 0.0% for every parameter, and 78 of 100 profiles recovered all five parameters within 5% of their true values. The remaining 22 profiles reflect a known structural-identifiability limitation rather than execution failure: under slow-absorption conditions, $K_a$ and $V_c$ are individually poorly determined, whereas their product is recovered accurately (Supplementary Figure 3).

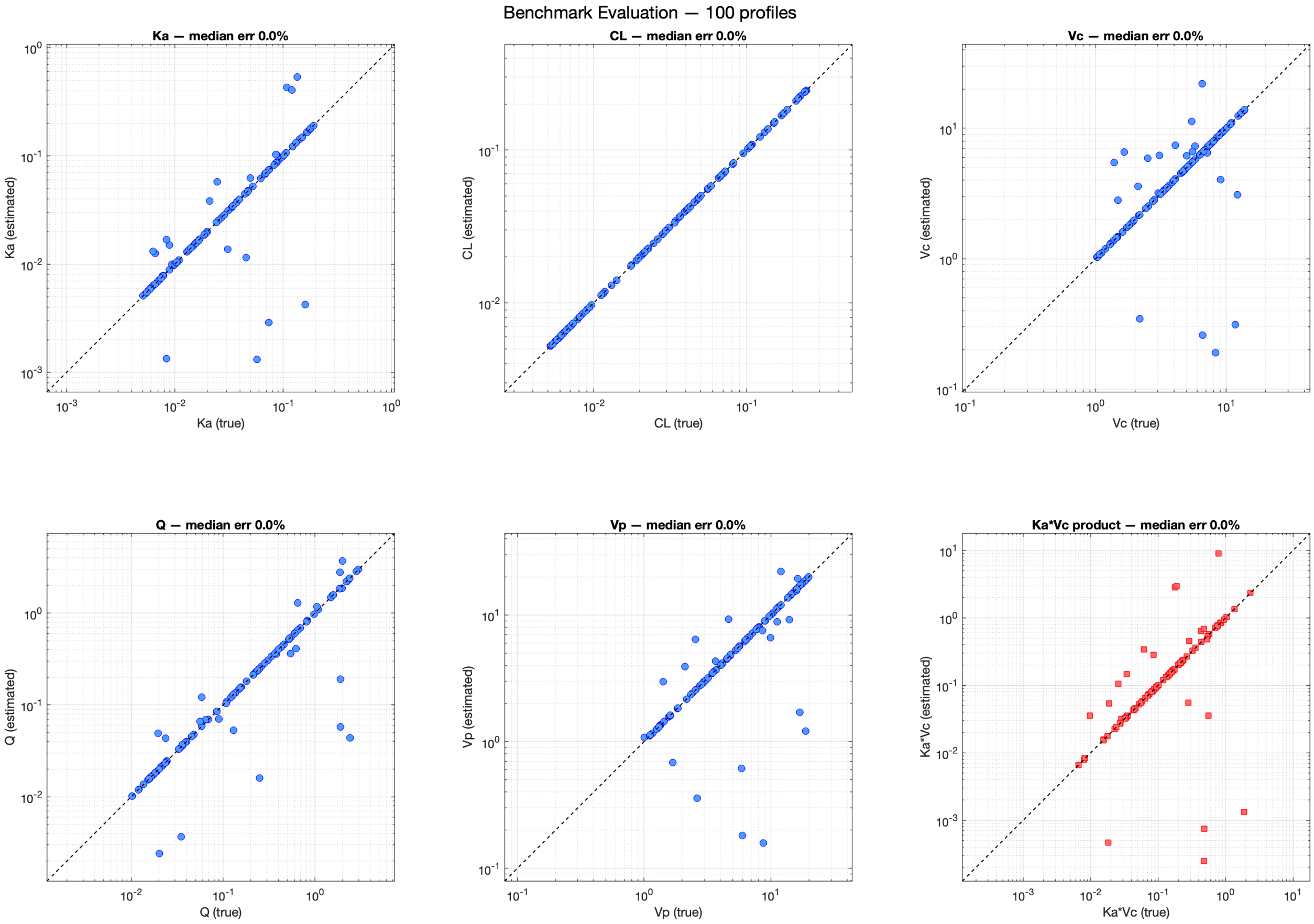

**Supplementary Figure 3:** Long-horizon autonomous PK fitting. Estimated versus ground-truth values for the five pharmacokinetic parameters ($K_a$, CL, $V_c$, Q, $V_p$) and the Ka·Vc product across 100 synthetic two-compartment profiles fitted autonomously by the MAS. Median error is 0.0% for every parameter; the near-perfect recovery of the $K_a \cdot V_c$ product indicates that the residual per-parameter scatter in $K_a$ and $V_c$ arises due to known structural identifiability limitations.

## *Head-to-head against a single-pass baseline*

To isolate the contribution of the multi-agent architecture, we compared the MAS with a single-pass GHCP baseline using `Claude 4.6 Opus`, the same frontier LLM used as the Optimization Agent's writer. We retained the 20 most difficult cases from the 100 executed PK profiles—those graded D or F under the identifiability scheme described in Methods—which represented the most severe fitting failures. Supplementary Figure 4 compares estimated and ground-truth values for the MAS and GHCP. The MAS recovered all parameters within 10% of their true values for 70% of profiles, compared with 20% for GHCP, and achieved median errors below 2% for every parameter ($K_a$, 0.2%; CL, 0.0%; $V_c$, 0.1%; Q, 1.5%; $V_p$, 0.5%). GHCP matched the MAS only for clearance (CL, 0.2%) and produced substantially larger median errors for $K_a$ (45%), $V_c$ (45%), Q (47%), and $V_p$ (33%), the parameters most affected by flip-flop kinetics and inter-compartmental non-identifiability.

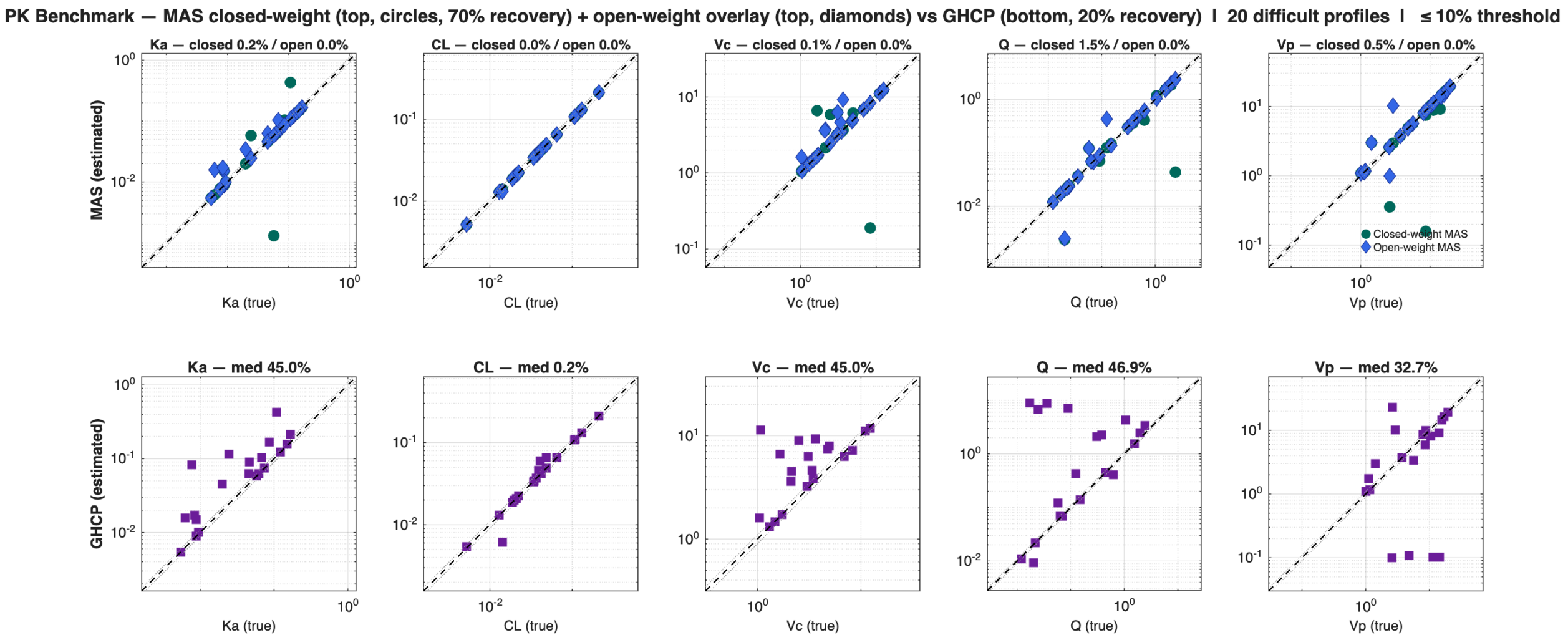

**Supplementary Figure 4:** MAS versus single-pass baseline (GHCP) on the twenty hardest PK profiles (graded D/F). Estimated (y-axis) versus ground-truth (x-axis) values for the MAS (top row) and GHCP (bottom row).

# Supplementary Note 4. Robustness of parameter recovery to the writer-agent LLM

To test whether MAS performance depends on the underlying writer LLM, we repeated the PKPD benchmark on the 11 most structurally challenging datasets using two open-weight writer configurations and compared them with the proprietary frontier-writer configuration. The tested systems used `DeepSeek-V4-Flash` or `DeepSeek-V4-Pro` for the writer role and `Llama 3.1 70B` for code review; the frontier comparison used `Claude Opus 4.8` as the writer. All three configurations selected the correct PK and PD structures for all 11 datasets. `DeepSeek-V4-Flash` closely matched `Claude Opus 4.8` in parameter recovery, with median absolute parameter errors of 6.8% and 7.0%, mean errors of 15.7% and 16.2%, and maximum errors of 92.4% and 90.9%, respectively, across 77 parameter comparisons. `DeepSeek-V4-Pro` retained exact structural selection but produced a higher median error of 9.0% and an extreme $E_{max}$ outlier that increased its mean and maximum errors to 2,465.0% and 140,522.3%, respectively. These results show that the architectural workflow is compatible with open-weight writers, while also demonstrating that nominally higher model capability does not guarantee better parameter recovery within a stochastic optimization workflow.

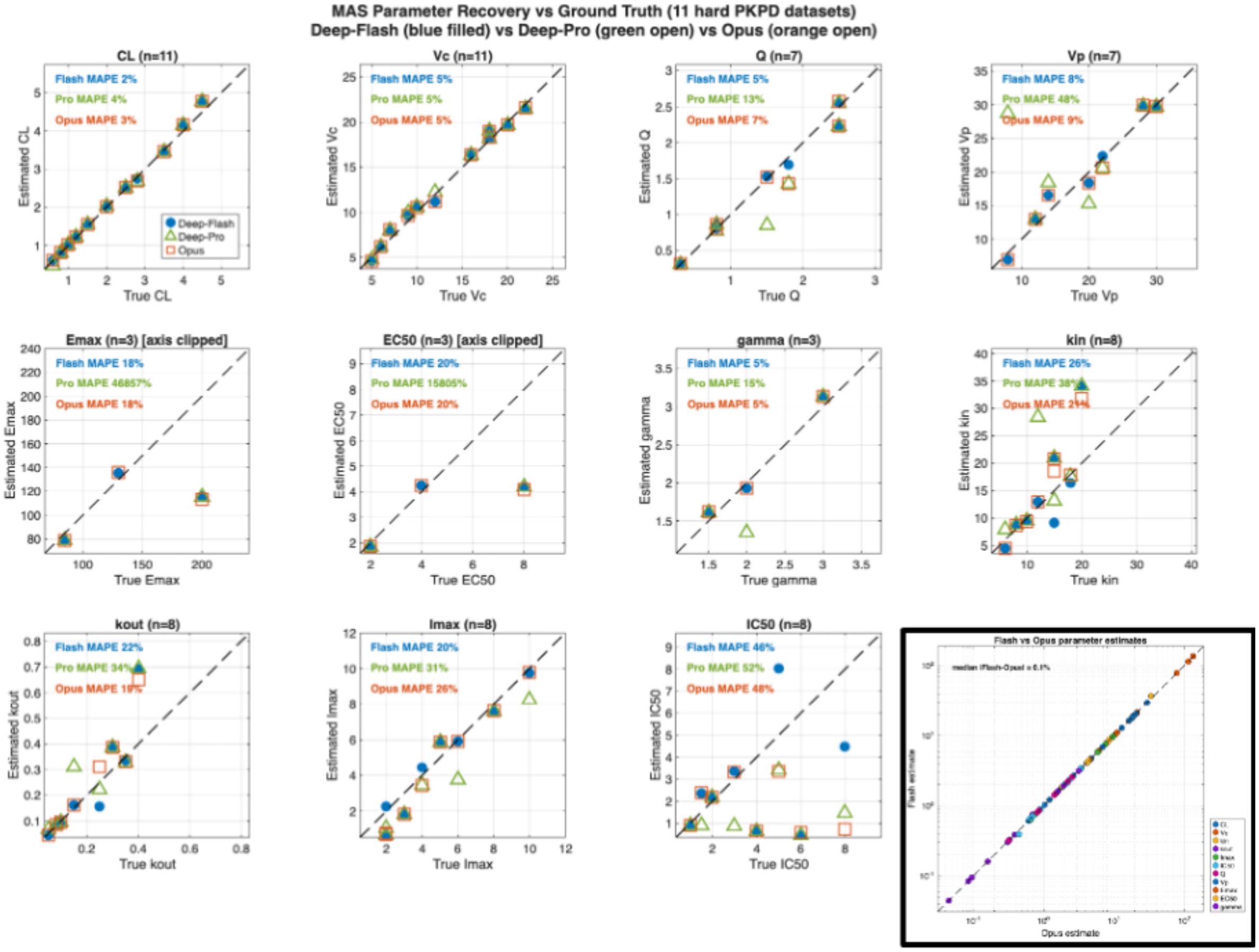

**Supplementary Figure 5. Parameter recovery across writer-model configurations.** Estimated parameter values are plotted against ground-truth values for the 11 most structurally challenging PKPD datasets using `DeepSeek-V4-Flash`, `DeepSeek-V4-Pro`, and `Claude Opus 4.8` as writer models. All three configurations achieved exact PK and PD structure selection for all 11 datasets (11/11). The $E_{max}$ and $EC_{50}$ axes are truncated to the ground-truth range to preserve visibility because `DeepSeek-V4-Pro` produced an extreme $E_{max}$ outlier for Dataset 14; the reported `DeepSeek-V4-Pro` mean absolute percentage error includes the full, untruncated value. The inset compares parameter recovery for `DeepSeek-V4-Flash` and `Claude Opus 4.8`, which differed by 0.1 percentage point and showed nearly identical performance.

# Supplementary Note 5. Model Implementation Workflow and System Prompts

To enable autonomous reproduction of complex mathematical models directly from published literature, the MAS follows a strictly constrained implementation pipeline (below). The Infrastructure Agent (Sam) first text-extracts the source paper, then calls the `analyze_paper_gaps` function to determine whether the paper has missing equations, equations residing in supplementary materials, or an associated public code repository (e.g. GitHub or Zenodo). When a public repository is available, Sam retrieves the reference implementation directly through the repository's API; otherwise, Sam extracts the governing equations from the paper using LLM vision and stores them, together with parameter units, initial conditions, and table footnotes, in a structured YAML file (`equations.yaml`). After the Modeling Agent (Alex) implements the ODE system in MATLAB, Alex runs a post-hoc symbolic verification (`verify_ode_equations.py`) that compares the MATLAB right-hand sides against the reference equations to confirm structural fidelity. The complete master prompt governing this PDF-to-code reproduction pipeline is provided below.

```
PUBLISHED MODEL IMPLEMENTATION WORKFLOW

When implementing a model from a published paper, ALWAYS follow this sequence:

1. Sam downloads & extracts the paper → PDF to pub_papers/, text to .md

2. Sam runs analyze_paper_gaps(paper_path) → identifies missing equations, supplementary refs, code repos

3. You review the gap report before delegating implementation

4a. If a GitHub/Zenodo repo is identified → Sam fetches reference code via API:

   fetch_webpage(url="https://api.github.com/repos/USER/REPO/contents/")

   Then navigates subdirs and fetches files with the correct branch (master vs main).

   ⚠️ DO NOT guess raw.githubusercontent.com paths — always list the repo first!

   📁 Sam MUST save each fetched file to: [active_project_folder]/ref_code/<paper_key>/filename

   Example: ref_code/pilvankar_2019/call_PKPD_model_scalar.m

   This enables automatic reference-code injection during PI review.

4b. If NO code repo is found → Sam extracts equations using vision:

   analyze_image(path="pub_papers/paper.pdf", page=N,

     question="Extract ALL ODE equations on this page in YAML format for symbolic verification. Use
EXPLICIT multiplication operators (*). Format:

     ```yaml

     model: 'Model Name'

     source: 'Author et al., Journal Year, Equations N-M'

     states:

       X: {{index: 1, name: 'Description'}}

     parameters:

       beta: {{name: 'Rate constant', units: 'day^-1'}}

     equations:

       dX_dt: 'expression with explicit * operators'

     equation_labels:

       dX_dt: 'Eq. 1'

     initial_conditions:
```
```

````
       X: 'X0'

     ```

     Include ALL state variables, parameters with units, and initial conditions.

     ⚠️ ALSO extract every TABLE FOOTNOTE that qualifies a parameter value — especially

     absorption/infusion PHASE TIME WINDOWS (e.g. 'k01: t = 0-120 min', 'k02: t = 120-300 min'),

     dose/bioavailability qualifiers, and any 'assumed 0 per GoF'-style notes. Record these

     verbatim as a `footnotes:` block in equations.yaml — do NOT paraphrase or round time

     windows from memory/context; transcribe the exact numbers printed in the table.")

   Sam saves extracted equations to equations.yaml (NOT LaTeX markdown).

   Repeat for each page with equations (especially older/scanned papers where text extraction garbles
math).

4c. MANDATORY SELF-VERIFICATION before implementation begins (do NOT skip this step):

   Delegate to Sam or Alex: "Re-read pub_papers/<paper>.md (or the PDF pages containing the

   parameter table) side-by-side with the just-generated equations.yaml. For every

   time-windowed or piecewise parameter (absorption phases, infusion durations, dosing

   intervals), confirm the equations.yaml values/windows match the table text EXACTLY.

   Then perform a back-of-envelope MASS-BALANCE / DOSE-CONSERVATION check: integrate the

   input/absorption rate(s) over their stated time windows and confirm the total delivered

   mass is consistent with the administered dose (within ~10%, accounting for bioavailability

   F). Report any discrepancy found — do not silently proceed if the numbers don't reconcile."

   This step exists because equation-extraction agents (including frontier models) reliably

   catch these errors WHEN EXPLICITLY ASKED to cross-check + compute mass balance, but will

   NOT catch them if merely asked to "implement from equations.yaml" — implementation fidelity

   to a flawed YAML is not the same as fidelity to the paper.

5. Alex/Maya implement with paper text AND fetched reference files OR extracted equations

6. After Alex writes the MATLAB ODE file, run verification:

   run_python("tools/verify_ode_equations.py <matlab_ode_file> <equations.yaml>")
````
````

## *Open-weight models for low- and high-complexity model implementation*

The main-text ablation study uses the low-complexity atenolol minimal physiologically based pharmacokinetic model of Kir et al. [9] to isolate the contributions of individual architecture components under a fixed implementation task. Here, we instead evaluated whether the complete architecture could reproduce published pharmacokinetic models spanning markedly different levels of structural complexity using open-weight writer models. We first examined reproduction of the four-state atenolol model and then evaluated implementation of the substantially larger 93-state monoclonal-antibody physiologically based pharmacokinetic model of Shah and Betts [2].

For the low-complexity task, `DeepSeek-V4-Flash` reproduced Figure 4 of Kir et al. [9] across the intravenous, oral control, and oral malnourished scenarios (Supplementary Fig. 6a). The workflow extracted the governing equations and parameter values from the provided publication, implemented the four-state ODE system in MATLAB, and generated forward simulations for comparison with manually digitized reference profiles. This result demonstrates that the structured literature-to-code pipeline can support faithful implementation of a compact PBPK model using a lower-tier open-weight writer end-to-end.

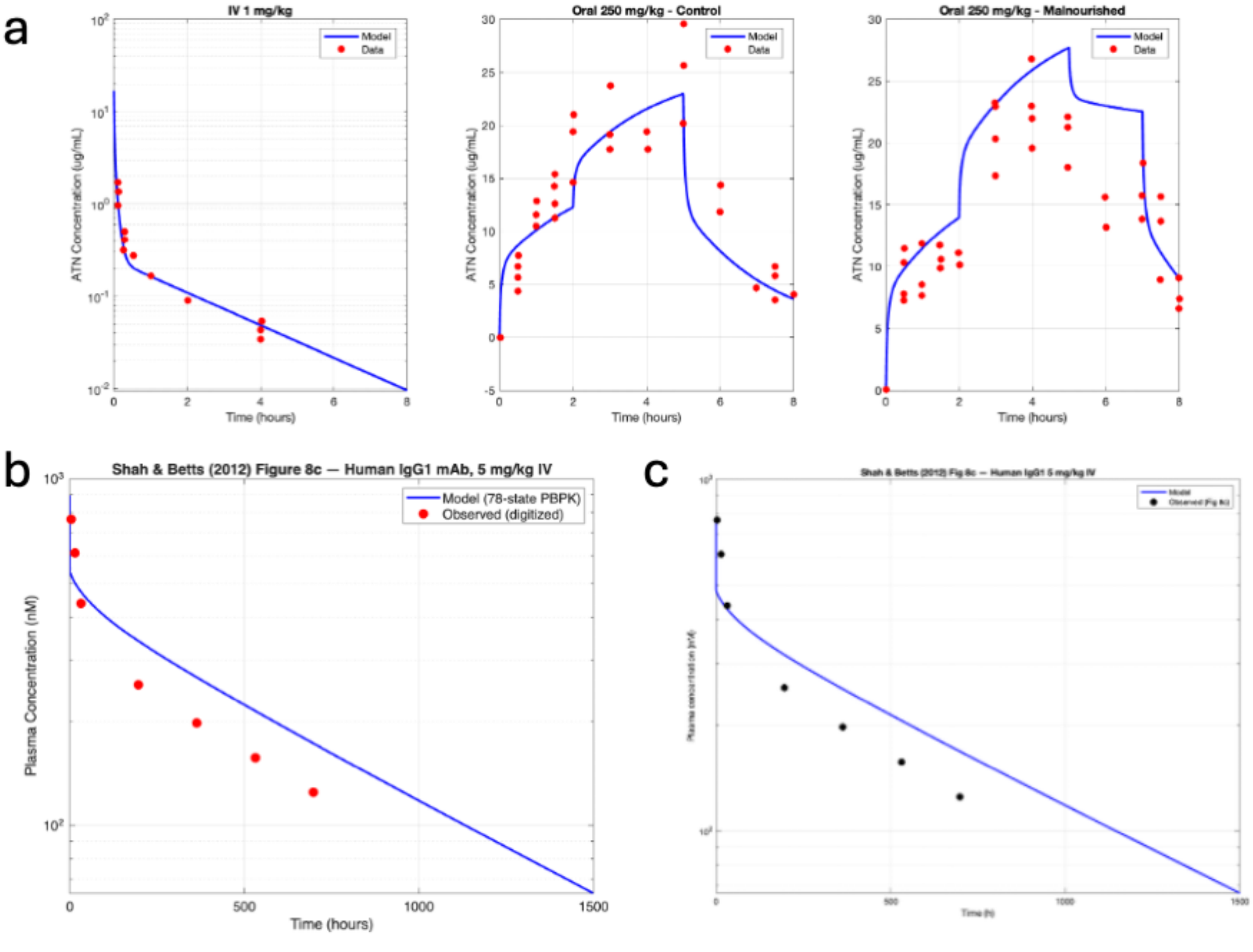

**Supplementary Figure 6.** Open-weight reproduction of published PBPK models spanning low and high structural complexity**. a**, Reproduction of Figure 4 of Kir et al. [9] using `DeepSeek-V4-Flash` as the writer agent, showing simulations for the intravenous, oral control, and oral

malnourished atenolol scenarios. **b**, Reproduction of Figure 8 of Shah and Betts [2] using `DeepSeek-V4-Pro` as the writer agent for the 93-state monoclonal-antibody PBPK model. **c**, Reproduction of Figure 8 of Shah and Betts [2] using `DeepSeek-V4-Flash` as the writer agent for the 93-state monoclonal-antibody PBPK model.

For the high-complexity task, `DeepSeek-V4-Pro` successfully reproduced Figure 8 of the Shah and Betts model (Supplementary Fig. 6b). This implementation required translation and verification of a 93-state system with tissue-specific transport, FcRn-mediated recycling, and explicit mass-balance constraints. In three independent attempts using `DeepSeek-V4-Flash`, the system diagnosed and corrected the mass-balance inconsistency sufficiently to reproduce Figure 8 in two runs but did not resolve the issue in the third. Thus, mass-balance diagnosis was achievable with the lower-tier open-weight writer but was not fully reliable across replicates. This variability identifies conservation-law verification as a practical bottleneck for open-weight models as model dimensionality and inter-compartmental coupling increase, while showing that a more capable open-weight writer can complete the same workflow.

Together, these examples demonstrate that the architecture supports literature-based model reproduction with open-weight models across a broad complexity range, although writer capability and run-to-run reliability remain consequential for high-dimensional systems with stringent physical constraints. `DeepSeek-V4-Flash` was sufficient for the four-state atenolol model and reproduced Figure 8 of the 93-state Shah–Betts ODE model in two of three independent attempts after diagnosing and correcting the mass-balance issue. `DeepSeek-V4-Pro` produced the successful high-complexity reproduction shown in Supplementary Figure 6b. The architecture also remains compatible with proprietary models, including Claude and Gemini models, which produced successful implementations in separate runs not shown here. Thus, the framework is not dependent on proprietary frontier models, but reliable physical-consistency verification in a complex computational model may require a more capable open-weight writer.

# Supplementary Note 6. Linguistic Invariance Test

To assess whether the PI Agent normalizes linguistic variation among users issuing the same scientific task, we presented Dataset 11—generated from a two-compartment PK model with direct $E_{max}$ PD—using nine distinct styles. Every prompt retained the same six candidate structures, multi-start optimization requirement, AIC-based selection rule, and output schema; only tone, register, grammar, formatting, and language varied. The PI Agent correctly parsed and delegated every prompt. The initial combined run nevertheless produced seven exact structural matches, one PK-structure miss (terse imperative), and one selection of the nested sigmoidal $E_{max}$ model (stream of consciousness). The passive-academic run selected the correct structure but returned severely mis-scaled PK estimates. Identical reruns were therefore performed for the terse-imperative and passive-academic prompts to characterize stochastic optimization reliability rather than linguistic comprehension. The prompts are reproduced verbatim below, followed by the original and rerun outcomes.

## *Terse imperative: A prompt containing minimal words with a commanding style*

```
Dataset 11. data/pkpd_benchmark_data.mat. Fit all 6 PKPD structures (1cmt/2cmt × Emax/SigEmax/IDR).

Multi-start ≥5. Select by AIC.

CSV: results/pkpd_benchmark_results.csv. Columns: dataset_id, selected_pk_model, selected_pd_model,
CL_est, Vc_est, Q_est, Vp_est, Emax_est, EC50_est, gamma_est, kin_est, kout_est, Imax_est, IC50_est,

best_WSSE, best_AIC, n_models_fitted, unique_solution, notes.

Plot: results/fit_dataset_11_overlay.png.
```

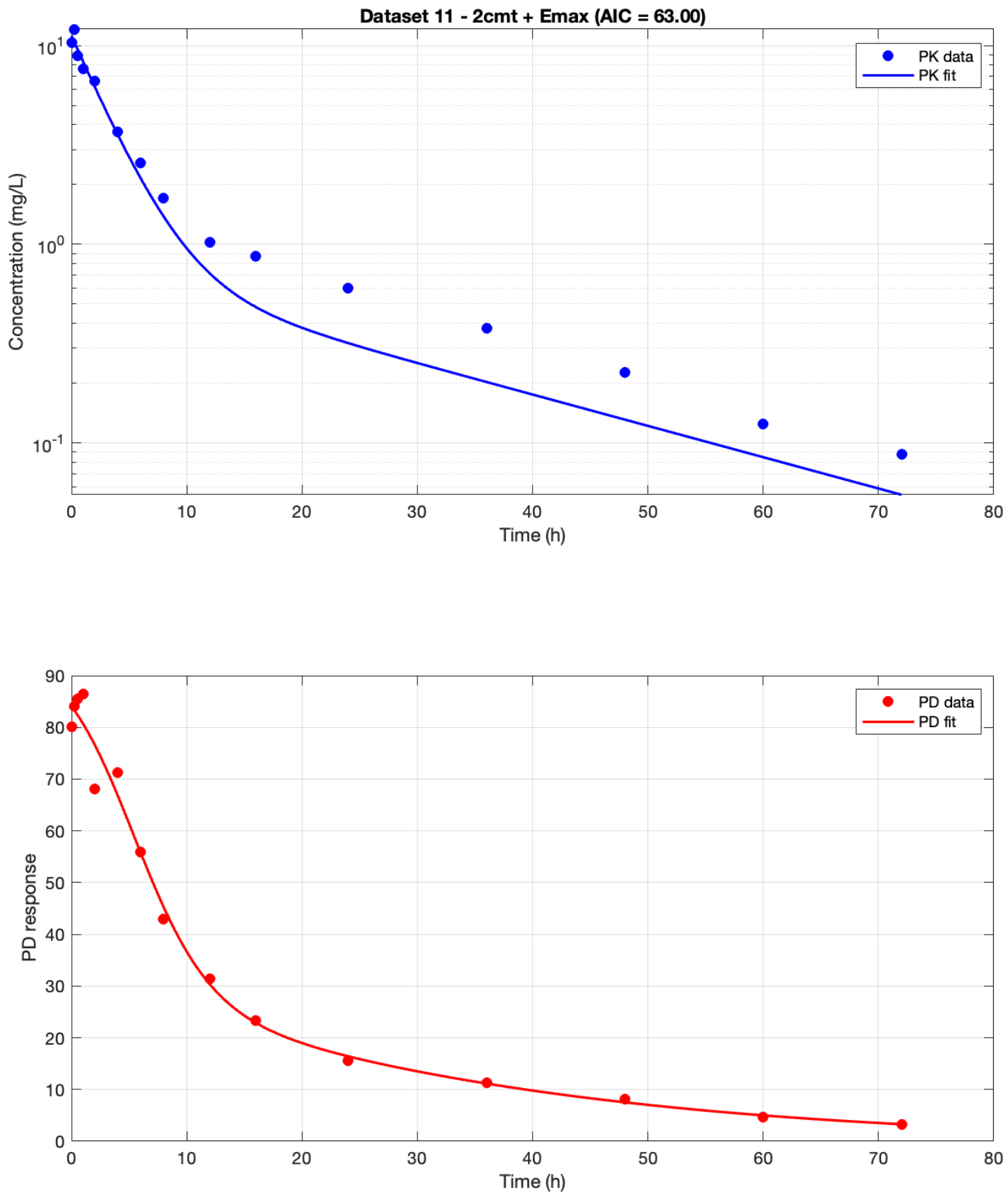

**Supplementary Figure 7.** Terse-imperative prompt. The initial combined run selected one-compartment PK plus direct $E_{max}$ PD. Repeating the identical prompt recovered the ground-truth two-compartment PK plus direct $E_{max}$ PD structure, with mean, median, and maximum absolute parameter errors of 20.1%, 16.5%, and 48.9%, respectively, across six parameter comparisons.

# *Formal British: A polite hedging prompt with passive constructions*

```
Good afternoon. I should be most grateful if you would kindly undertake a model selection exercise on
dataset number 11 from the benchmark data file located at data/pkpd_benchmark_data.mat.



The candidate model structures to be evaluated are the six combinations arising from two pharmacokinetic
alternatives (one-compartment and two-compartment disposition) crossed with three pharmacodynamic
alternatives (direct Emax response, sigmoidal Emax response, and indirect response via inhibition of
elimination).



It would be appreciated if a multi-start optimisation procedure were employed, with no fewer than five
random starting parameter vectors per model, so as to mitigate the risk of convergence to local minima.
Model selection ought to be conducted on the basis of the Akaike Information Criterion.



The results should be written in CSV format to results/pkpd_benchmark_results.csv, with columns:
dataset_id, selected_pk_model, selected_pd_model, CL_est, Vc_est, Q_est, Vp_est,

Emax_est, EC50_est, gamma_est, kin_est, kout_est, Imax_est, IC50_est, best_WSSE, best_AIC,
n_models_fitted, unique_solution, notes.



An overlay plot of the observed and fitted PK and PD profiles would also be most helpful, saved as
results/fit_dataset_11_overlay.png. Many thanks indeed.
```

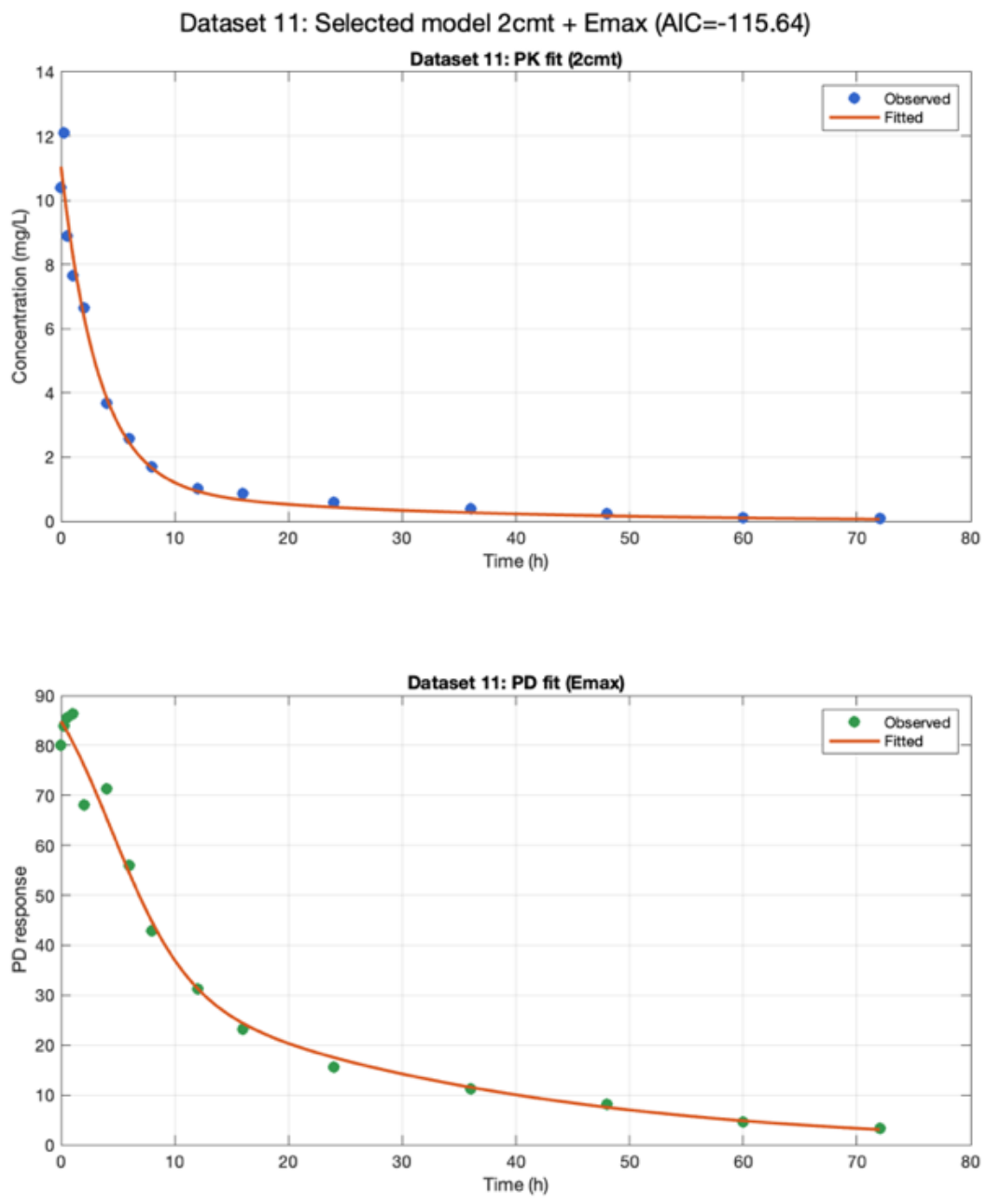

**Supplementary Figure 8.** Formal-British prompt result. The run selected the ground-truth two-compartment PK plus direct $E_{max}$ PD structure, with a mean absolute parameter error of 13.4% across six parameter comparisons.

## *Casual American: A conversational prompt with contractions*

```
Hey! So I've got dataset 11 sitting in data/pkpd_benchmark_data.mat and I need you to run all six PKPD
model combos on it — that's the 1-compartment and 2-compartment PK paired up with Emax, sigmoidal Emax,
and indirect response (IDR) for the PD side. Make sure you do multi-start optimization with at least 5
random starts per model so we don't get stuck in local minima, and then pick the winner using AIC.



Dump the results into results/pkpd_benchmark_results.csv with these columns:

dataset_id, selected_pk_model, selected_pd_model, CL_est, Vc_est, Q_est, Vp_est, Emax_est, EC50_est,
gamma_est, kin_est, kout_est, Imax_est, IC50_est, best_WSSE, best_AIC, n_models_fitted, unique_solution,
notes.



Oh, and throw in an overlay plot at results/fit_dataset_11_overlay.png showing the data vs the best fit.
Thanks!
```

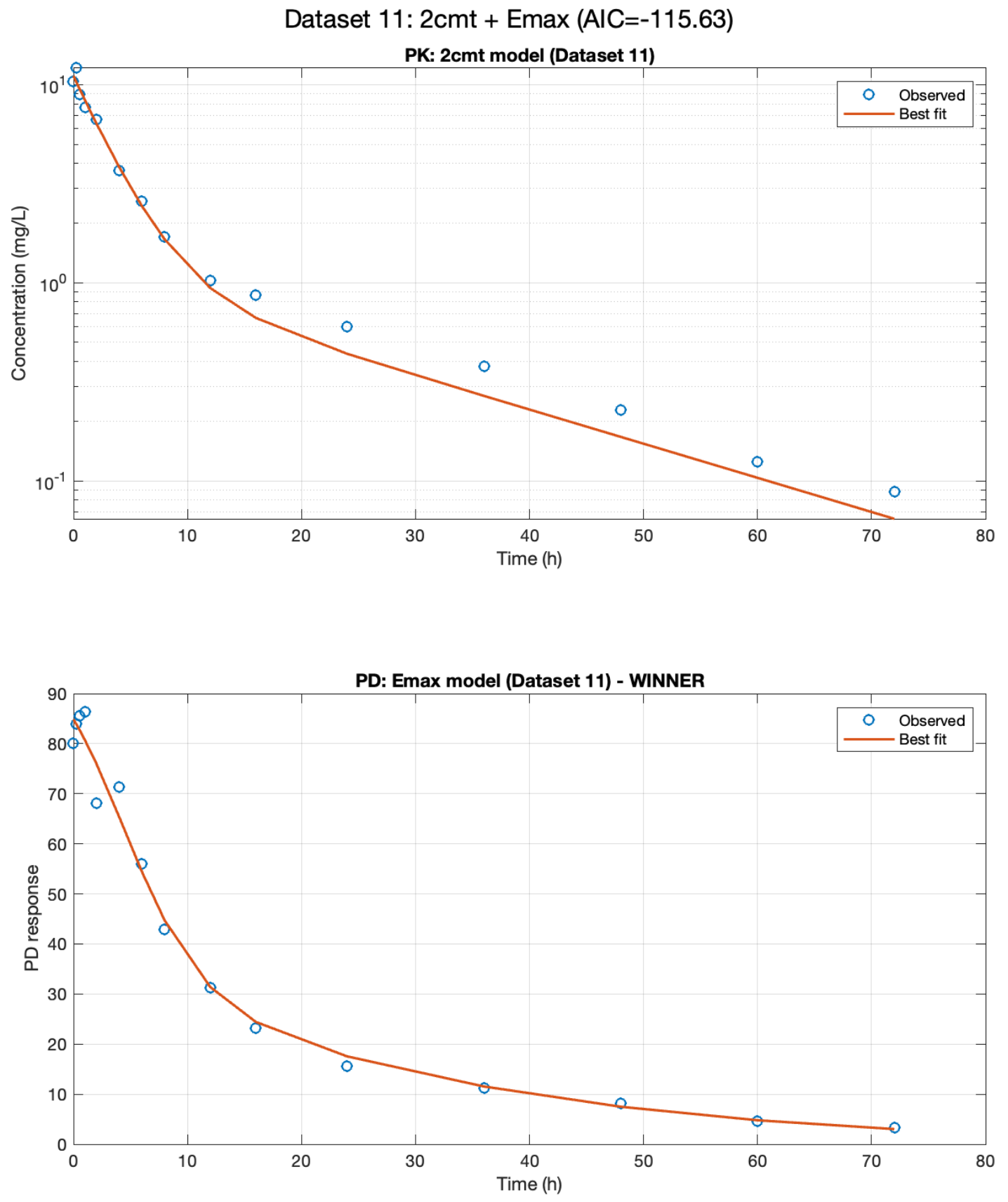

**Supplementary Figure 9.** Casual-American prompt result. The run selected the ground-truth two-compartment PK plus direct $E_{max}$ PD structure, with a mean absolute parameter error of 13.5% across six parameter comparisons.

## *Broken grammar: A non-native speaker prompt with grammatical errors*

```
Hello, I am needing to do the fitting of dataset 11 from the file data/pkpd_benchmark_data.mat please.
Must to test all 6 PKPD model structure — this is meaning one-compartment and two-compartment for PK part,
and for each one testing with Emax model, with sigmoidal Emax model, and with indirect response model (the
inhibition type). Please to be using the multi-start optimization, minimum 5 start for each model, and
choosing the best one by AIC criteria.


Save the result to CSV file at results/pkpd_benchmark_results.csv please. The column must be: dataset_id,
selected_pk_model, selected_pd_model, CL_est, Vc_est, Q_est, Vp_est, Emax_est, EC50_est, gamma_est,
kin_est, kout_est, Imax_est, IC50_est, best_WSSE, best_AIC, n_models_fitted, unique_solution, notes.


Also I am needing overlay plot save to results/fit_dataset_11_overlay.png showing observed data and best
model fit. Thank you very much for helping.
```

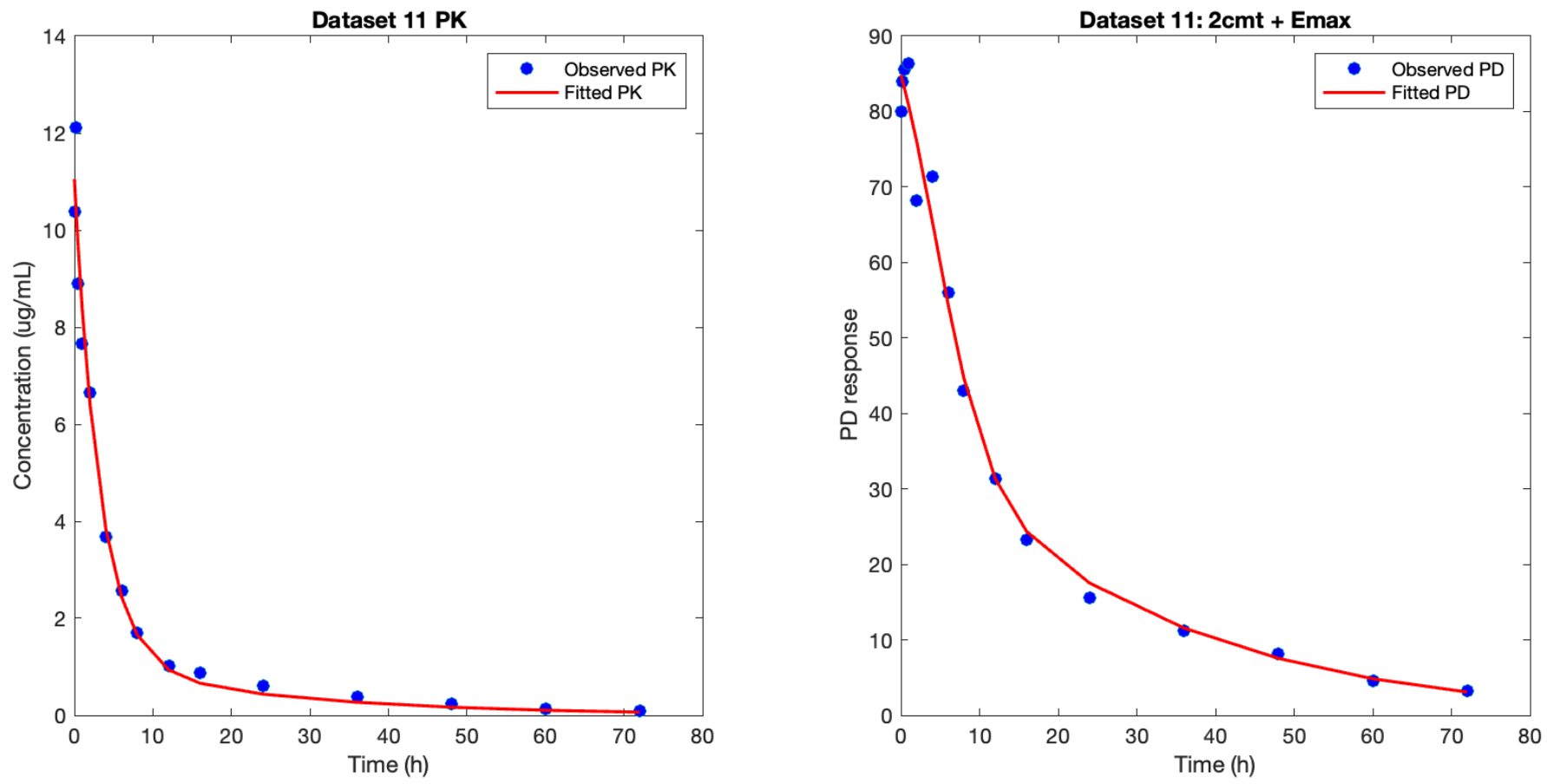

**Supplementary Figure 10.** Broken-grammar prompt result. The run selected the ground-truth two-compartment PK plus direct $E_{max}$ PD structure, with a mean absolute parameter error of 13.4% across six parameter comparisons.

## *Bullet points: A structured bullet-point format prompt with no prose*

```
Input: data/pkpd_benchmark_data.mat, dataset 11


Models to fit (6 total):

- 1-compartment PK + direct Emax PD

- 1-compartment PK + sigmoidal Emax PD

- 1-compartment PK + indirect response PD (inhibition)

- 2-compartment PK + direct Emax PD
```

```
- 2-compartment PK + sigmoidal Emax PD

- 2-compartment PK + indirect response PD (inhibition)



Optimization: multi-start, n_starts ≥ 5 per model



Selection criterion: AIC (lowest wins)



Output CSV: results/pkpd_benchmark_results.csv

- Columns: dataset_id, selected_pk_model, selected_pd_model, CL_est, Vc_est, Q_est, Vp_est, Emax_est,
EC50_est, gamma_est, kin_est, kout_est, Imax_est, IC50_est, best_WSSE, best_AIC, n_models_fitted,
unique_solution, notes



Output plot: results/fit_dataset_11_overlay.png (observed vs fitted overlay)
```

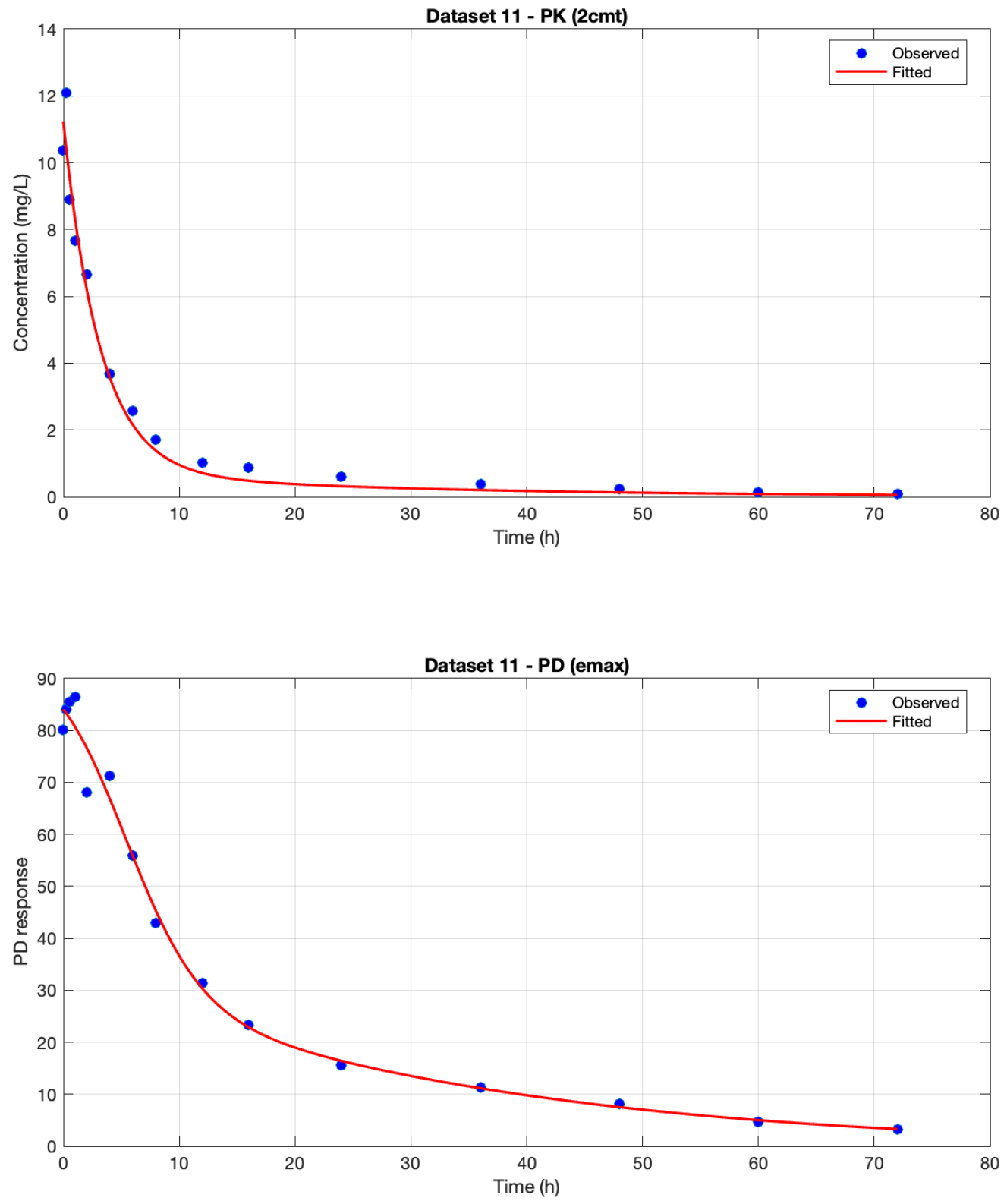

**Supplementary Figure 11.** Bullet-points prompt result. The run selected the ground-truth two-compartment PK plus direct $E_{max}$ PD structure, with a mean absolute parameter error of 20.1% across six parameter comparisons.

## *Stream of consciousness: A rambling and unorganized prompt*

```
OK so I was thinking about this — we have that benchmark data right, the pkpd_benchmark_data.mat file in
the data folder, and I specifically want to look at dataset 11. What I want to do is test all the PKPD
model combinations we normally use — so that's the six structures where you take the PK side

(one-compartment or two-compartment) and cross it with the PD side (Emax, sigmoidal Emax, or indirect
response with inhibition of elimination) — actually now that I think about it, make sure you use multi-
start optimization because last time someone just ran a single start and got trapped in a local

minimum, so let's do at least 5 random starting vectors per model to be safe.

Then just pick the best one by AIC as usual. For the output, I need the standard CSV at
results/pkpd_benchmark_results.csv — you know the format: dataset_id, selected_pk_model,
selected_pd_model, CL_est, Vc_est, Q_est, Vp_est, Emax_est, EC50_est, gamma_est, kin_est, kout_est,
Imax_est, IC50_est, best_WSSE, best_AIC, n_models_fitted, unique_solution, notes.

Oh and also generate an overlay plot showing the data points and the best model fit — save that to
results/fit_dataset_11_overlay.png. I think that covers everything.
```

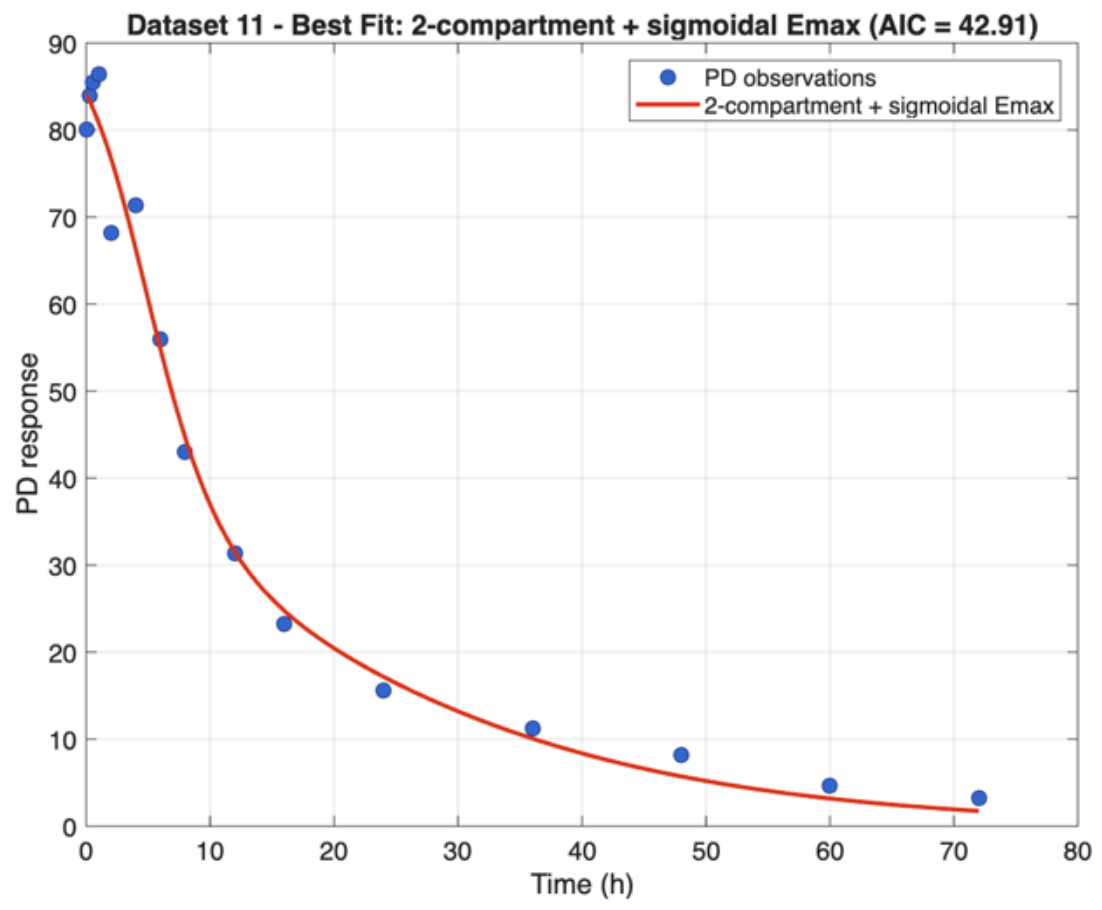

**Supplementary Figure 12.** Stream-of-consciousness prompt. The run selected two-compartment PK plus sigmoidal $E_{max}$ PD (AIC = 42.91; WSSE = 175.6) over the ground-truth two-compartment PK plus direct $E_{max}$ model (AIC = 45.42; WSSE = 237.2). The AIC difference of 2.51 modestly exceeded the Burnham–Anderson threshold for meaningful separation. The estimated sigmoidicity coefficient was approximately 1.2–1.3, close to the direct $E_{max}$ limit of $\gamma = 1$, indicating that the selected response shape remained close to the generating model.

## *Mixed language: A prompt switching between Japanese and English*

```
お願いします。Dataset 11 の fitting を実行してください。データファイルは

data/pkpd_benchmark_data.mat です。



Test all 6 PKPD model structures — つまり、1-compartment PK と

2-compartment PK の 2 種類に対して、それぞれ Emax、sigmoidal Emax、
```

```
indirect response (inhibition) の 3 種類の PD model を組み合わせた 6 通りです。


Optimization: multi-start (最低 5 回の random starting vectors)を使って、

AIC で best model を選択してください。


結果を CSV で保存: results/pkpd_benchmark_results.csv

Columns: dataset_id, selected_pk_model, selected_pd_model, CL_est, Vc_est,

Q_est, Vp_est, Emax_est, EC50_est, gamma_est, kin_est, kout_est, Imax_est,

IC50_est, best_WSSE, best_AIC, n_models_fitted, unique_solution, notes.


Overlay plot も作成してください: results/fit_dataset_11_overlay.png

ありがとうございます。
```

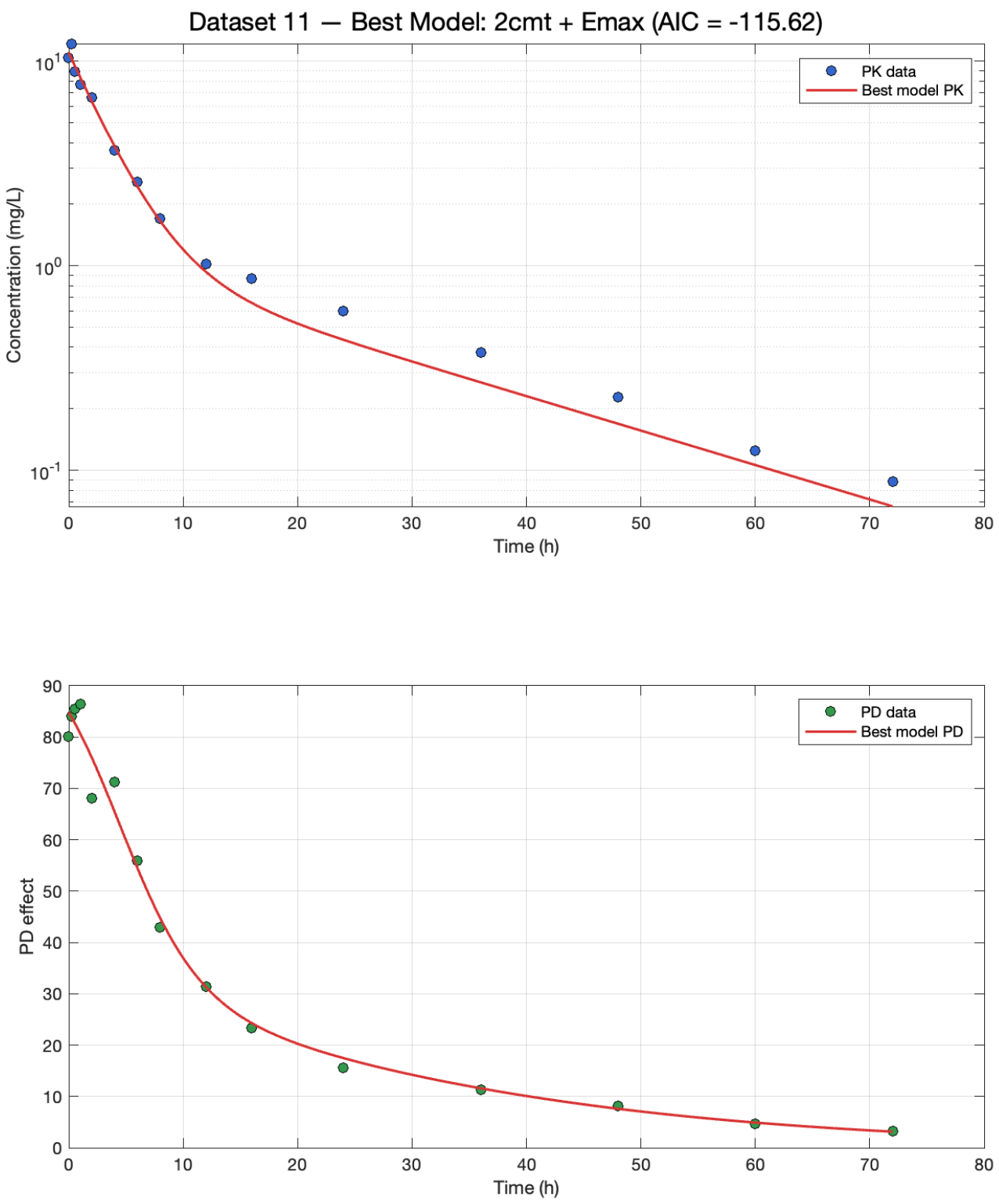

**Supplementary Figure 13.** Mixed-language prompt result. The run selected the ground-truth two-compartment PK plus direct $E_{max}$ PD structure, with a mean absolute parameter error of 13.3% across six parameter comparisons.

# *Passive academic: A nominalized academic prompt*

```
The following analysis is requested to be performed. The pharmacokinetic and pharmacodynamic data
contained within dataset 11 of the file located at data/pkpd_benchmark_data.mat are to be subjected to
model fitting.



A total of six candidate model structures are to be evaluated, comprising the factorial combination of two
pharmacokinetic structural alternatives (a one-compartment disposition model and a two-compartment
disposition model) with three pharmacodynamic structural alternatives (a direct Emax model, a

sigmoidal Emax model, and an indirect response model characterised by inhibition of the elimination
process).



Parameter estimation is to be conducted via a multi-start optimisation approach, wherein a minimum of five
stochastic initial parameter vectors are to be employed for each candidate structure, thereby reducing the
probability of convergence to local optima. Model discrimination is to be performed on the basis of the
Akaike Information Criterion.



The results of the analysis are to be exported in comma-separated value format to
results/pkpd_benchmark_results.csv, incorporating the following fields: dataset_id, selected_pk_model,
selected_pd_model, CL_est, Vc_est, Q_est, Vp_est, Emax_est, EC50_est, gamma_est, kin_est, kout_est,
Imax_est, IC50_est, best_WSSE, best_AIC, n_models_fitted, unique_solution, notes.



Additionally, a graphical overlay of observed versus model-predicted concentrations and pharmacodynamic
responses is to be generated and saved to results/fit_dataset_11_overlay.png.
```

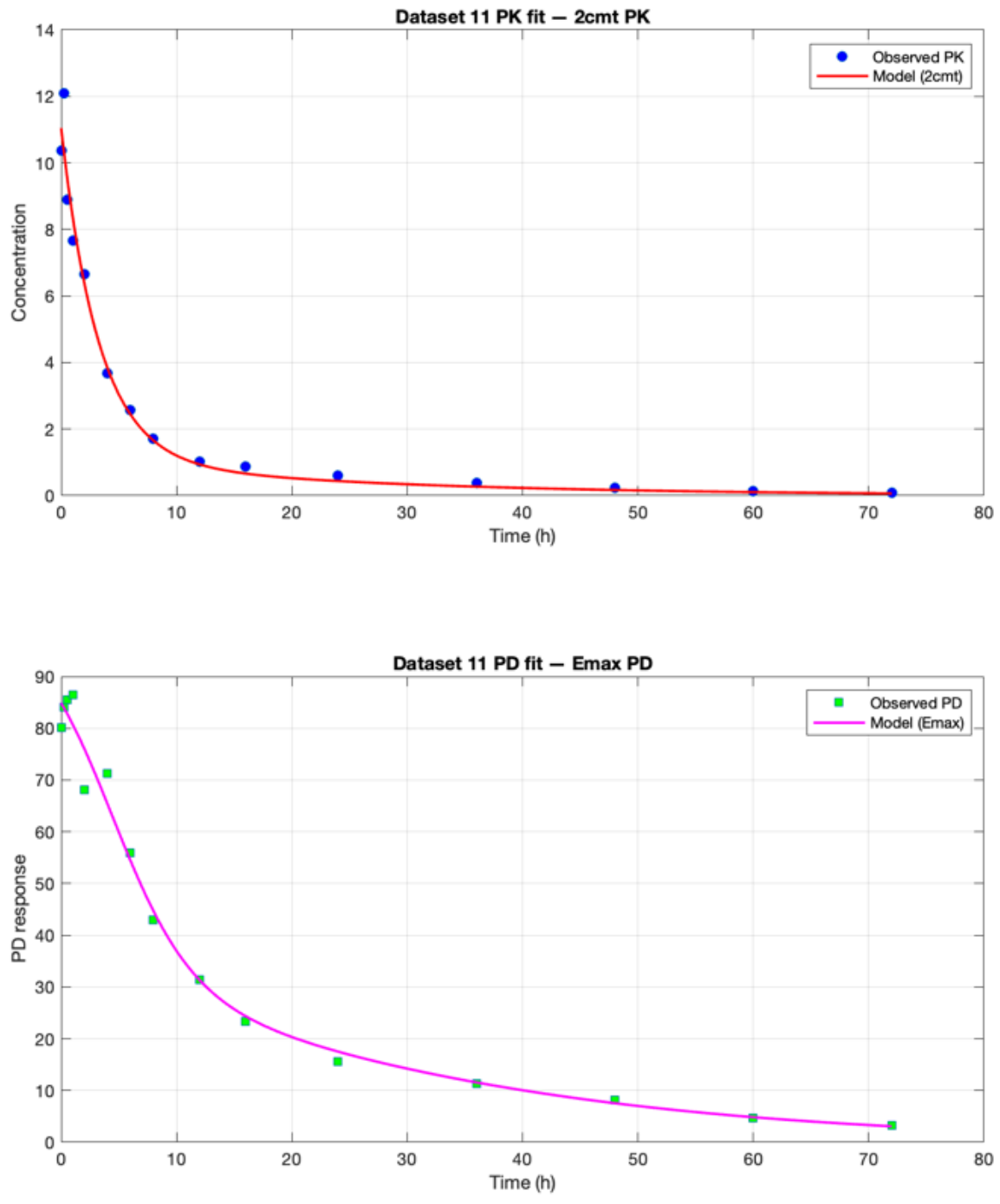

**Supplementary Figure 14.** Passive-academic prompt. The initial combined run selected the ground-truth two-compartment PK plus direct $E_{max}$ PD structure but produced an extreme mean parameter error of 63,151%, driven by implausibly scaled PK estimates. Repeating the identical

prompt recovered the same structure with mean, median, and maximum absolute parameter errors of 13.4%, 10.3%, and 30.7%, respectively, across six parameter comparisons.

## *Emoji slack: A chat style prompt with emojis*

```
🧪 hey team! need a model comparison run on DS11 👇

📂 data: data/pkpd_benchmark_data.mat (dataset 11)

🔬 models: all 6 PKPD combos

  - PK: 1cmt, 2cmt

  - PD: Emax, SigEmax, IDR (inhibition type)

⚙️ optimization: multi-start w/ ≥5 random starts per model

📊 selection: AIC

💾 output CSV → results/pkpd_benchmark_results.csv

cols: dataset_id, selected_pk_model, selected_pd_model, CL_est, Vc_est, \

Q_est, Vp_est, Emax_est, EC50_est, gamma_est, kin_est, kout_est, Imax_est, \

IC50_est, best_WSSE, best_AIC, n_models_fitted, unique_solution, notes

📈 overlay plot → results/fit_dataset_11_overlay.png (obs vs pred)

thx! 🙏
```

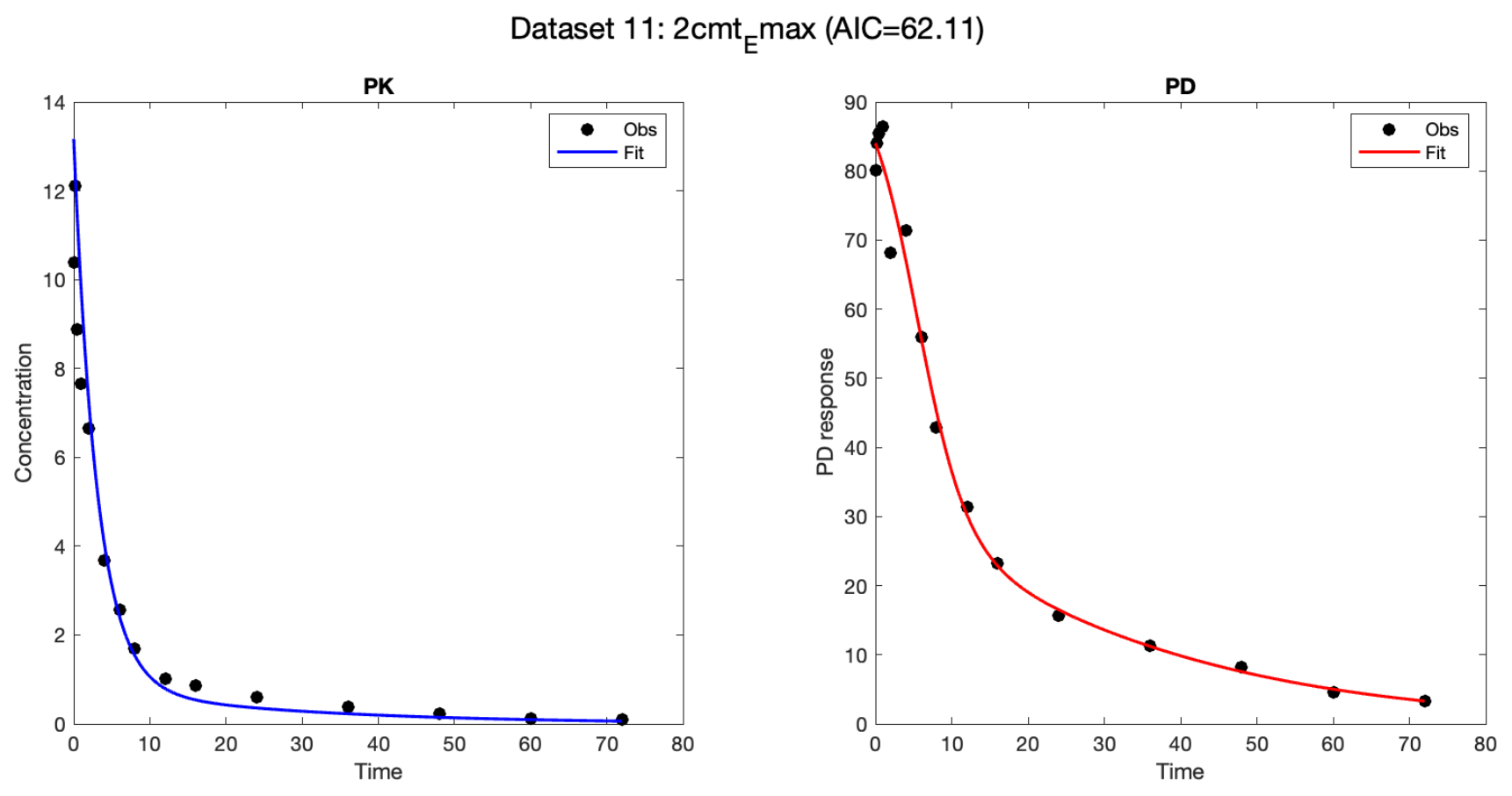

**Supplementary Figure 15.** Emoji-Slack prompt result. The run selected the ground-truth two-compartment PK plus direct $E_{max}$ PD structure, with a mean absolute parameter error of 22.9% across six parameter comparisons.

**Supplementary Table 1:** Mean parameter error across the nine linguistic variations†

| S. No. | Prompt flag | Initial structure | Final reported outcome | Mean error (%) |
|---|---|---|---|---|
| 1 | terse_imperative | 1cmt + Emax | 2cmt + Emax after identical rerun* | 20.1 |
| 2 | formal_british | 2cmt + Emax | Exact match | 13.4 |
| 3 | casual_american | 2cmt + Emax | Exact match | 13.5 |
| 4 | broken_grammar | 2cmt + Emax | Exact match | 13.4 |
| 5 | bullet_points | 2cmt + Emax | Exact match | 20.1 |
| 6 | stream_of_consciousness | 2cmt + sigEmax | Nested-model alternative; ΔAIC = 2.51; γ ≈ 1.3** | 14.2 |
| 7 | mixed_language | 2cmt + Emax | Exact match | 13.3 |
| 8 | passive_academic | 2cmt + Emax; severe scaling error | 2cmt + Emax after identical rerun* | 13.4 |
| 9 | emoji_slack | 2cmt + Emax | Exact match | 22.9 |

*Identical reruns are reported separately from the initial combined run. The terse-imperative rerun recovered the ground-truth structure with mean/median error of 20.1%/16.5%; the passive-academic rerun reduced the initial 63,151% mean error to 13.4% (median 10.3%). **For stream of consciousness, sigmoidal $E_{max}$ was favored over direct $E_{max}$ by ΔAIC = 2.51. Because $\gamma$ was approximately 1.3, the selected response shape was close to direct $E_{max}$ ($\gamma = 1$).

The initial combined run produced seven of nine exact structural matches. Terse imperative selected one-compartment rather than two-compartment PK, whereas stream of consciousness selected sigmoidal rather than direct $E_{max}$ PD. Passive academic selected the correct structure but produced an extreme mean parameter error of 63,151% because the PK estimates were severely mis-scaled. Identical reruns recovered the ground-truth structure for terse imperative and passive academic, with mean/median errors of 20.1%/16.5% and 13.4%/10.3%, respectively. The stream-of-consciousness alternative was scientifically close to the generating model: sigmoidal $E_{max}$ was favored by ΔAIC = 2.51, and $\gamma$ was approximately 1.2–1.3 versus $\gamma = 1$ for direct $E_{max}$. Accordingly, eight of nine variants recovered the ground-truth structure after reruns, and the ninth selected a closely related nested PD model. The key invariant outcome was correct task interpretation and execution across all nine linguistic forms; downstream optimization remained sensitive to stochastic convergence, scaling, and modest AIC differences between nested models.

# Supplementary Note 7. Prompt Robustness

The linguistic-invariance test held the task specification constant and varied only surface phrasing. In contrast, the prompt-robustness test varied the completeness and specificity of the task description while retaining the same scientific goal: model selection and parameter estimation for Dataset 11. The eight prompts ranged from a fully specified 1,383-character expert brief that enumerated all six candidate PKPD structures, the optimization procedure, selection criterion, and output schema to a 134-character instruction and a deliberately vague junior-researcher request. Independent rescoring of the archived outputs with `score_pkpd_benchmark.py` corrected false-negative classifications for the novice-vague and minimal variants, both of which selected the ground-truth two-compartment PK plus $E_{max}$ PD model. The expert-verbose prompt was rerun unchanged after its initial optimization selected an incorrect PK structure; the rerun recovered the ground-truth model. The eight prompts and corresponding results are provided below.

## *Expert concise: An experienced modeler prompt*

```
Fit dataset 11 from data/pkpd_benchmark_data.mat.

Test all 6 PKPD model structures (1cmt/2cmt × Emax/SigEmax/IDR).

Use multi-start optimization (≥5 starts). Select best model by AIC.

Report results to results/pkpd_benchmark_results.csv with columns:

dataset_id, selected_pk_model, selected_pd_model, CL_est, Vc_est, Q_est, Vp_est, Emax_est, EC50_est,
gamma_est, kin_est, kout_est, Imax_est, IC50_est, best_WSSE, best_AIC, n_models_fitted, unique_solution,
notes

Generate overlay plot: results/fit_dataset_11_overlay.png
```

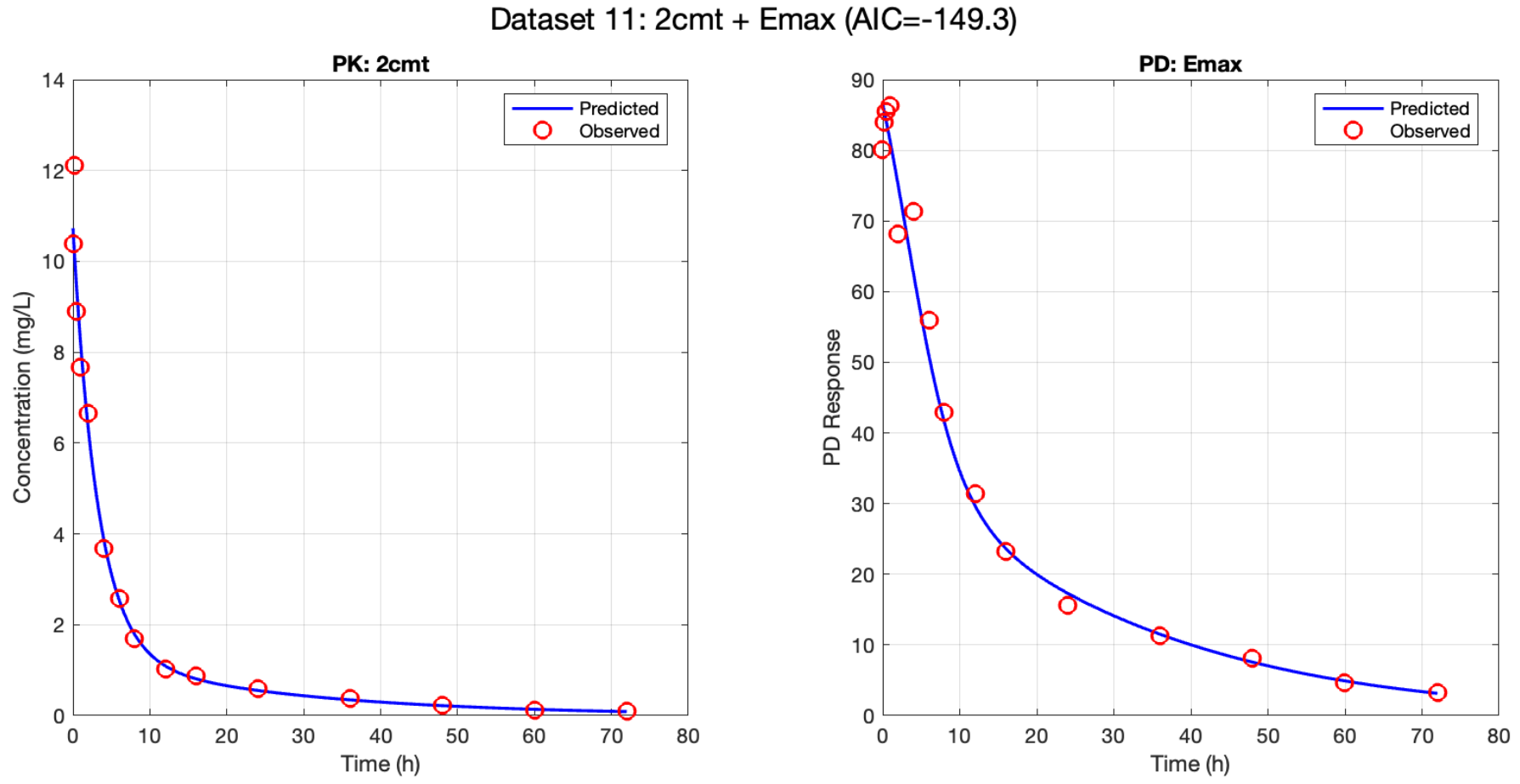

**Supplementary Figure 16.** Expert-concise prompt result. The run selected the ground-truth two-compartment PK plus direct $E_{max}$ PD structure.

## *Expert verbose: An experienced pharmacometrician*

```
Good morning. I'd like you to perform a model selection exercise on dataset number 11 from the benchmark
data file (data/pkpd_benchmark_data.mat).



The candidate model structures to evaluate are as follows:

- One-compartment PK with direct Emax pharmacodynamic response

- One-compartment PK with sigmoidal Emax response

- One-compartment PK with indirect response mechanism (inhibition of elimination)

- Two-compartment PK with direct Emax pharmacodynamic response

- Two-compartment PK with sigmoidal Emax response

- Two-compartment PK with indirect response mechanism



For each candidate model, please ensure you run a multi-start optimisation procedure with no fewer than 5
random starting parameter vectors to avoid local minima. Model selection should be performed using the
Akaike Information Criterion (AIC), and I would also like you to verify parameter identifiability (i.e.,
that the Hessian is positive-definite at the optimum).



Please save the results in CSV format to results/pkpd_benchmark_results.csv with these columns:
dataset_id, selected_pk_model, selected_pd_model, CL_est, Vc_est, Q_est, Vp_est, Emax_est, EC50_est,
gamma_est, kin_est, kout_est, Imax_est, IC50_est, best_WSSE, best_AIC, n_models_fitted, unique_solution,
notes.



Also produce an overlay plot showing observed vs. fitted PK and PD profiles saved as
results/fit_dataset_11_overlay.png. Many thanks.
```

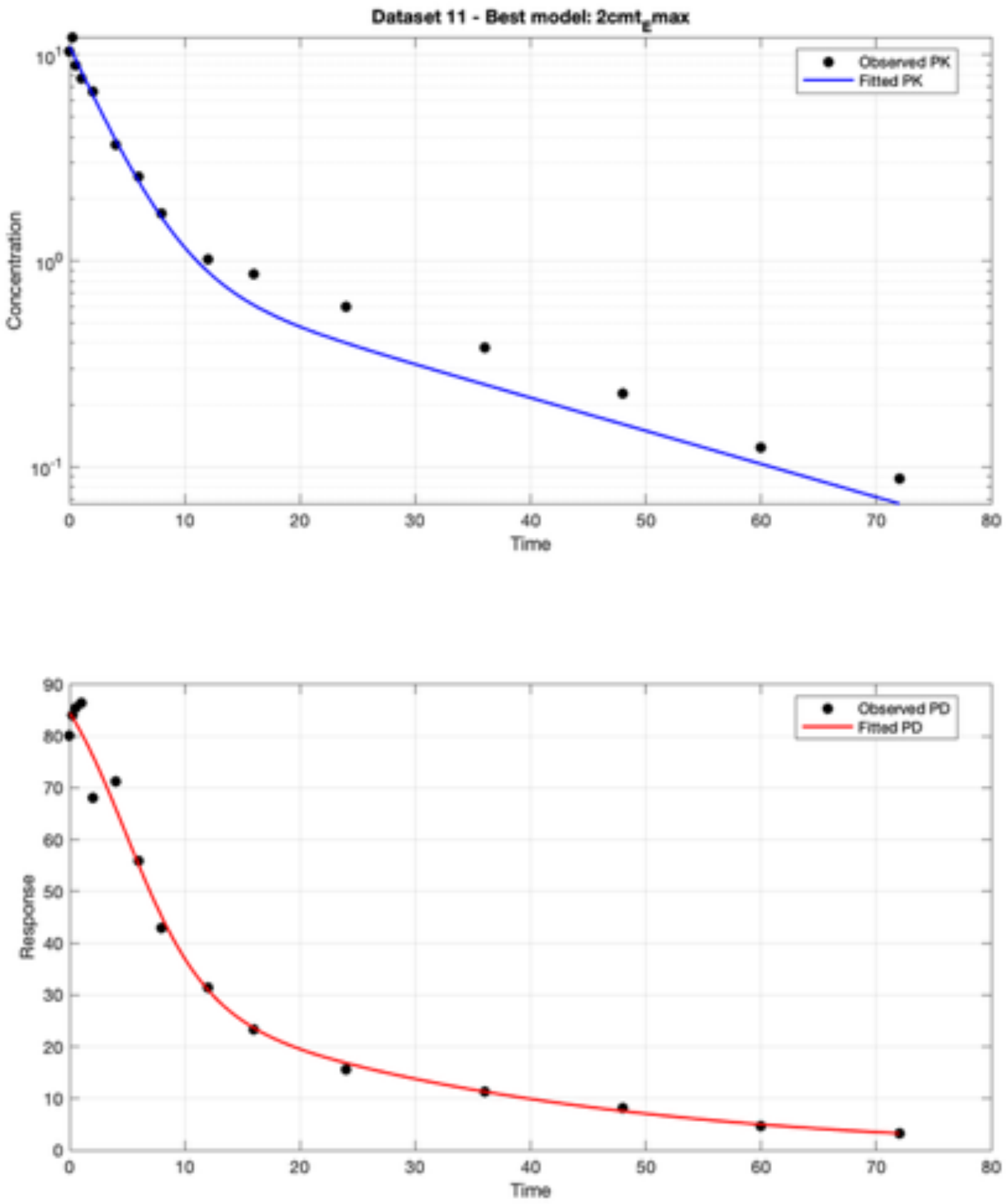

**Supplementary Figure 17.** Expert-verbose prompt rerun. The identical prompt recovered the ground-truth two-compartment PK plus $E_{max}$ PD model, with mean, median, and maximum absolute parameter errors of 15.0%, 10.8%, and 34.4%, respectively, across six parameter comparisons.

## *Novice vague: A vague prompt with uncertain terminology*

```
Hi, I have some PK/PD data in the mat file (data/pkpd_benchmark_data.mat) and
I need help figuring out which model fits best for dataset 11. I think it might
be a two-compartment model but I'm not sure? Could you try different models
and tell me which one works best? Also can you make a plot showing how well
it fits? I'd like the results saved somewhere I can find them, maybe a CSV
file in the results folder. Thanks!
```

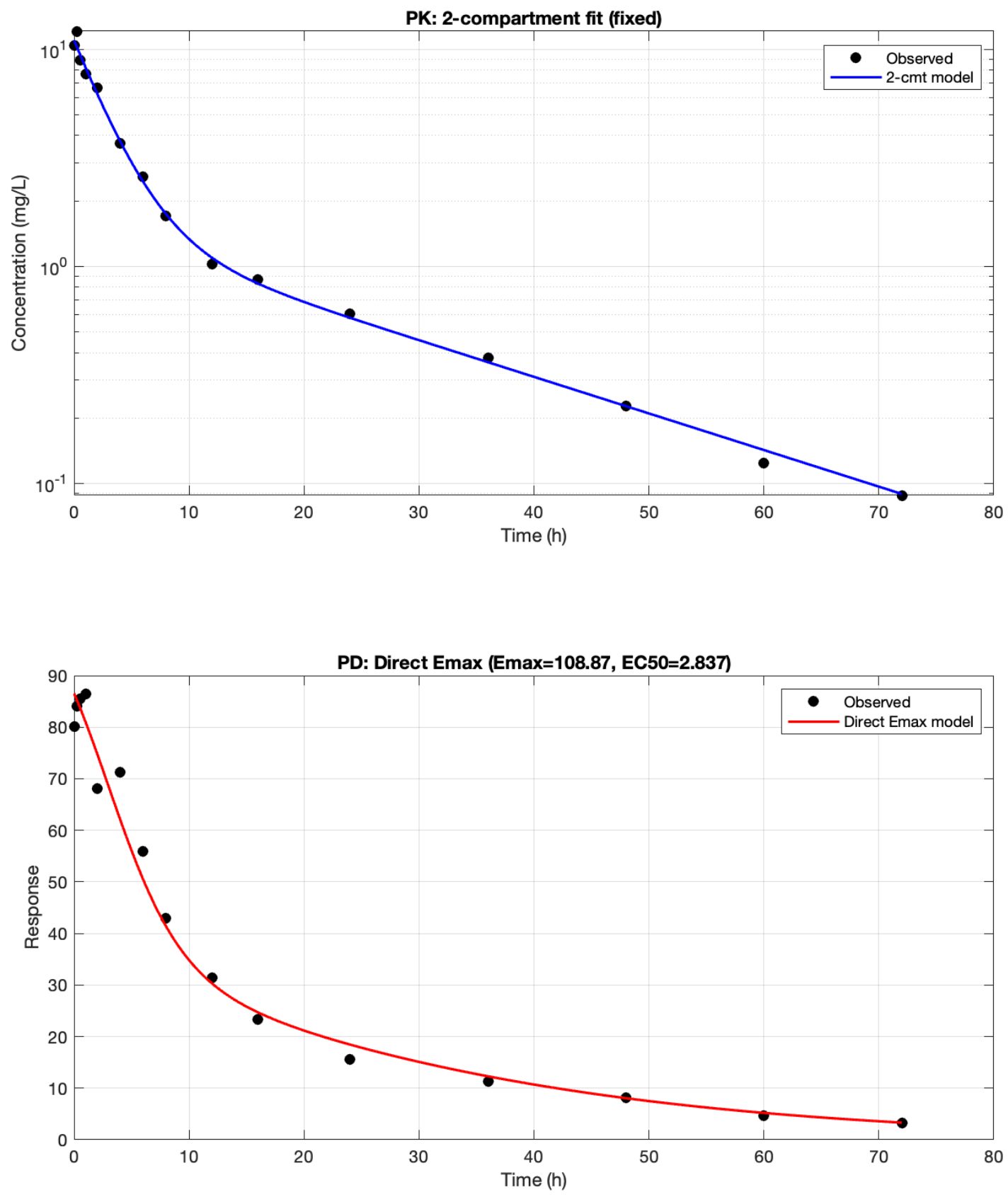

**Supplementary Figure 18.** Novice-vague prompt result. Independent rescoring confirmed selection of the ground-truth two-compartment PK plus $E_{max}$ PD model, with mean, median, and maximum absolute parameter errors of 5.4%, 3.2%, and 16.1%, respectively, across six parameter comparisons.

## *Non-native English prompt*

```
Please to be fitting dataset number 11 from the PKPD benchmark data (data/pkpd_benchmark_data.mat). I am
requiring to test all 6 model structures – one compartment and two compartment PK, each with Emax,
sigmoidal Emax, and indirect response PD models. Kindly use multi-start optimization with minimum 5 starts
and select best model using AIC criterion.

Results should be saved in CSV format at results/pkpd_benchmark_results.csv.

Please also generate fitting overlay plot at results/fit_dataset_11_overlay.png.

Column names for CSV: dataset_id, selected_pk_model, selected_pd_model, CL_est, Vc_est, Q_est, Vp_est,
Emax_est, EC50_est, gamma_est, kin_est, kout_est, Imax_est, IC50_est, best_WSSE, best_AIC,
n_models_fitted, unique_solution, notes. Thanking you.
```

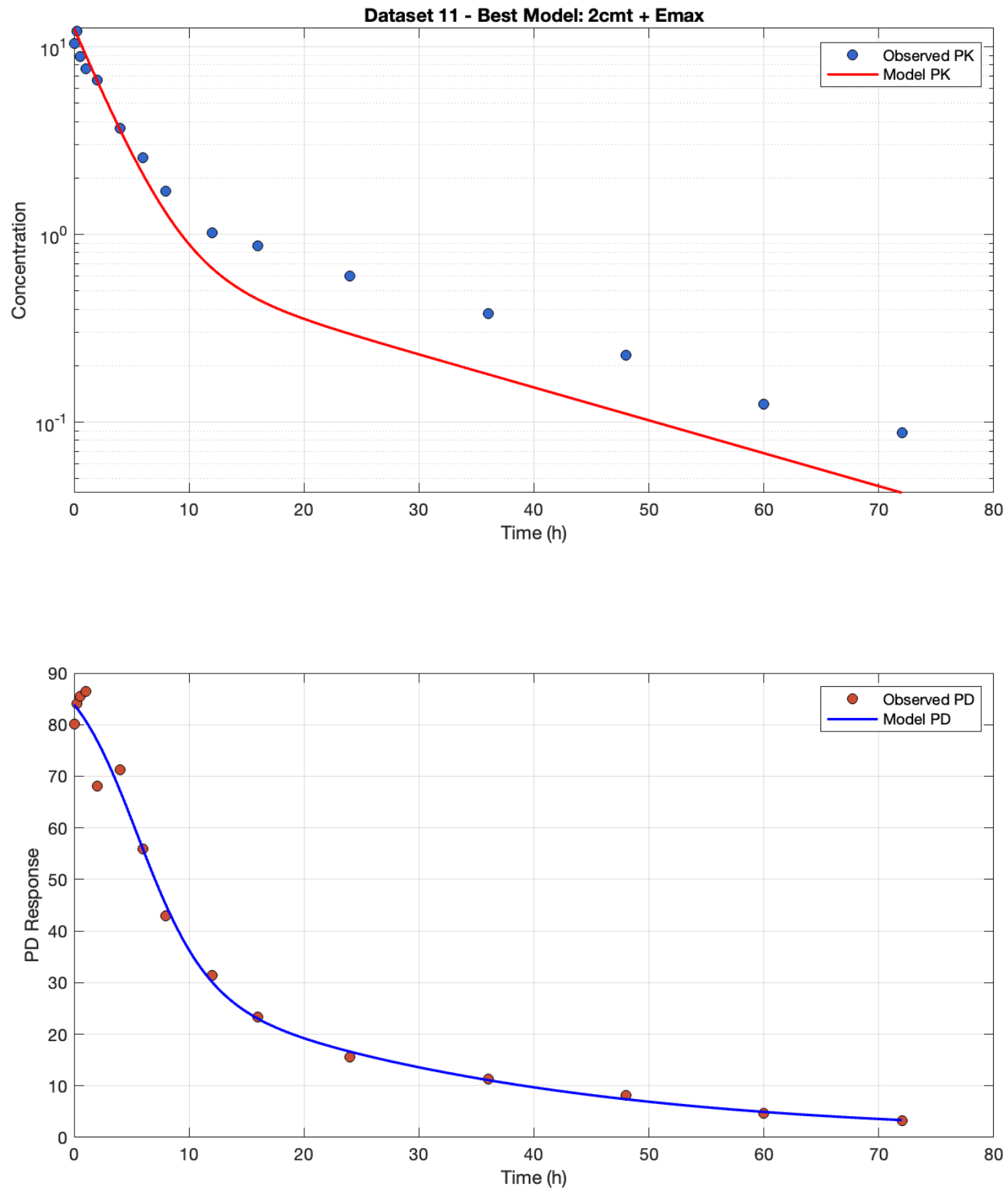

**Supplementary Figure 19.** Non-native-English prompt result. The run selected the ground-truth two-compartment PK plus direct $E_{max}$ PD structure.

## *Scandinavian formal: A scientist with formal and structured prompt*

```
Hej. I would like to request a structured model comparison for dataset 11 in data/pkpd_benchmark_data.mat.


Task specification:

• Candidate models: 6 structures (2 PK × 3 PD)

• PK alternatives: one-compartment, two-compartment

• PD alternatives: Emax, sigmoidal Emax, indirect response (inhibition)

• Fitting procedure: multi-start (n ≥ 5), weighted sum of squared errors

• Selection criterion: AIC

• Output: CSV to results/pkpd_benchmark_results.csv
```

```
• Visualization: overlay of observed and fitted profiles →

results/fit_dataset_11_overlay.png



CSV columns required: dataset_id, selected_pk_model, selected_pd_model, CL_est, Vc_est, Q_est, Vp_est,
Emax_est, EC50_est, gamma_est, kin_est, kout_est, Imax_est, IC50_est, best_WSSE, best_AIC,
n_models_fitted, unique_solution, notes.



Tack.
```

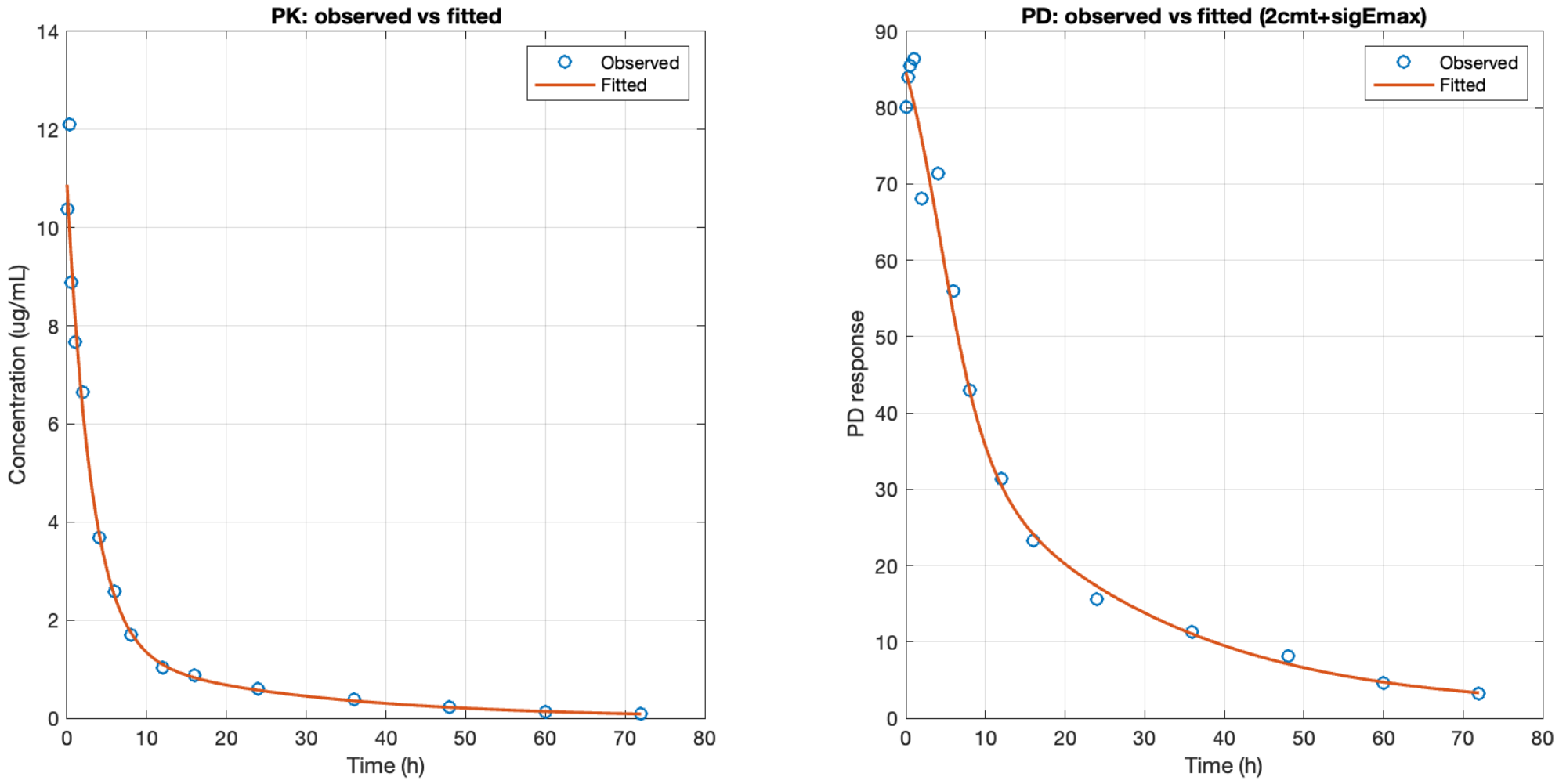

**Supplementary Figure 20.** Model-selection results for the Scandinavian-formal prompt. The two-compartment sigmoidal $E_{max}$ model achieved a slightly lower weighted sum of squared errors and AIC (WSSE = 0.1367; AIC = −145.7; eight parameters; 10/10 converged starts) than the two-compartment $E_{max}$ model (WSSE = 0.1548; AIC = −144.0; seven parameters; 10/10 converged starts). The estimated sigmoidicity coefficient was 1.2. Because the AIC difference was 1.7, the two models were not meaningfully distinguishable under the Burnham–Anderson criterion ($\Delta$AIC < 2).

## *Minimal prompt*

```
Fit dataset 11. All 6 models. Best by AIC. Save CSV to results/pkpd_benchmark_results.csv, plot to results/fit_dataset_11_overlay.png.
```

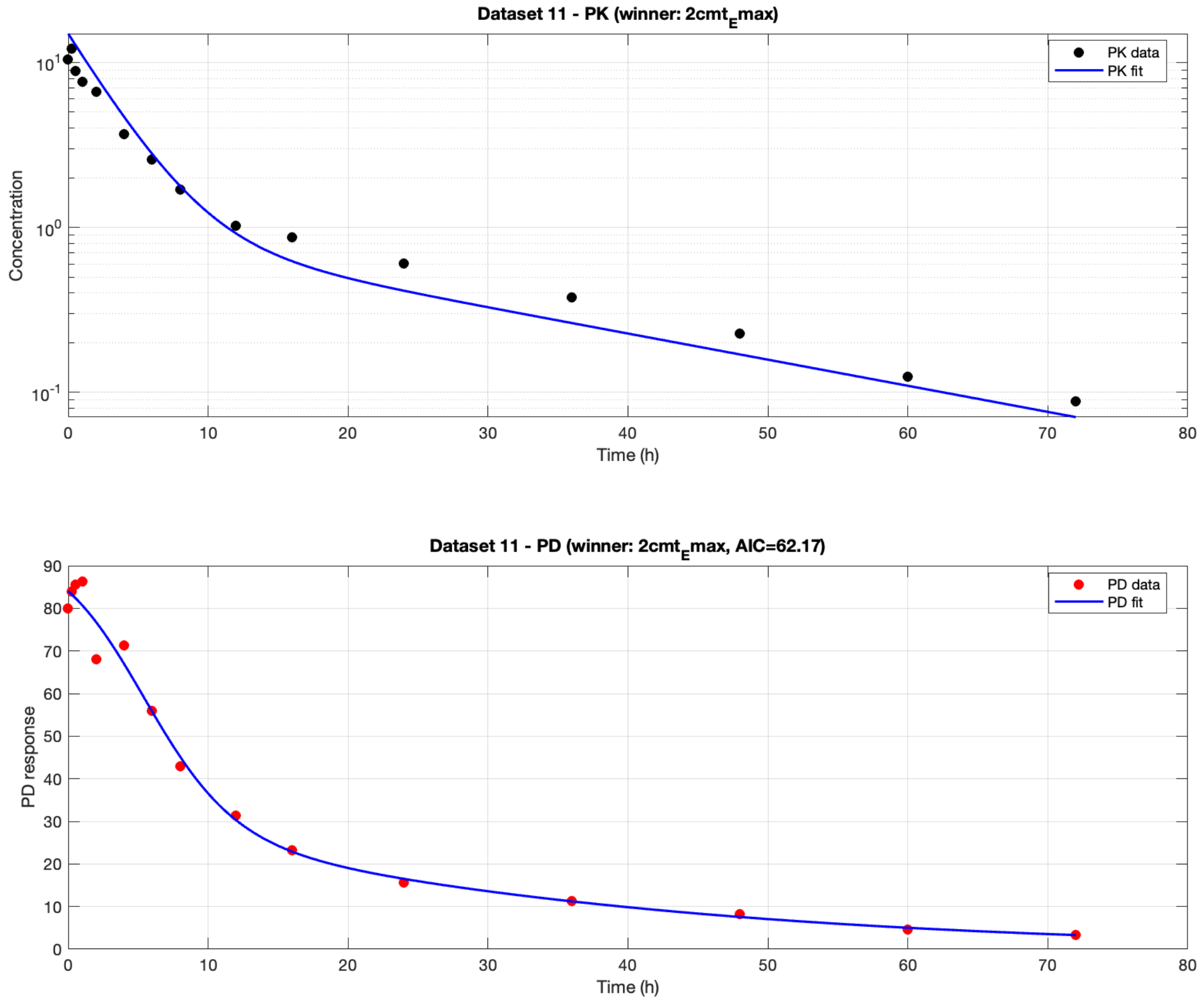

**Supplementary Figure 21.** Minimal-prompt result. Independent rescoring confirmed selection of the ground-truth two-compartment PK plus $E_{max}$ PD model, with mean, median, and maximum absolute parameter errors of 25.1%, 29.6%, and 38.9%, respectively, across six parameter comparisons.

# *Canadian collaborative: A scientist with polite and collaborative prompt*

```
Hey there, would you mind taking a look at dataset 11 in data/pkpd_benchmark_data.mat for me? I'd really
appreciate it if you could run all six candidate PKPD models — that's the one-compartment and two-
compartment PK options, each paired with Emax, sigmoidal Emax, and indirect response PD. If you could do
multi-start optimization (at least 5 starts, eh?) and pick the best model by AIC, that would be great.

Save the results as a CSV in results/pkpd_benchmark_results.csv with the usual columns: dataset_id,
selected_pk_model, selected_pd_model, CL_est, Vc_est, Q_est, Vp_est, Emax_est, EC50_est, gamma_est,
kin_est, kout_est, Imax_est, IC50_est, best_WSSE, best_AIC, n_models_fitted, unique_solution,

notes. And an overlay plot at results/fit_dataset_11_overlay.png would be perfect. Merci beaucoup!
```

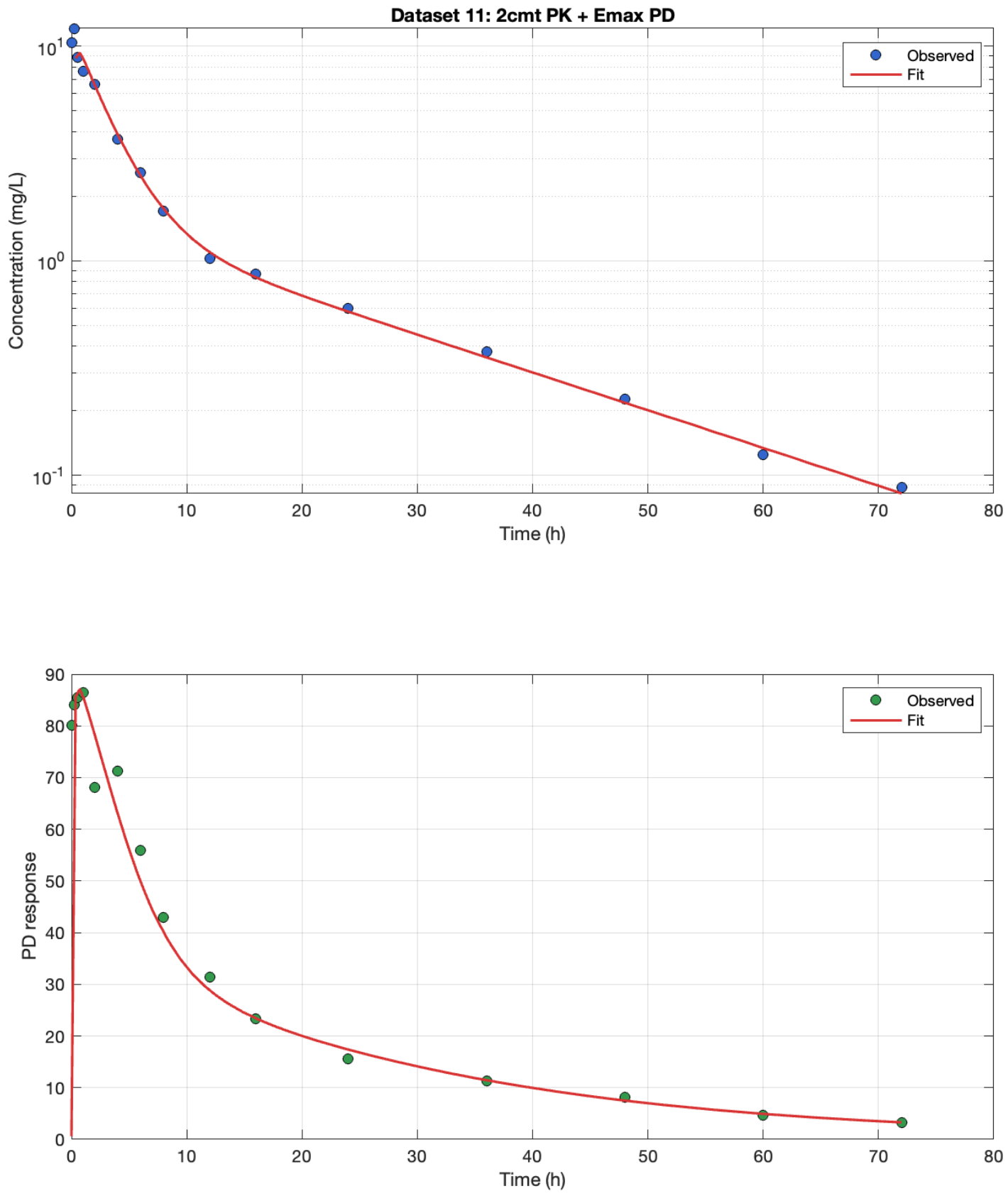

**Supplementary Figure 22.** Canadian-collaborative prompt result. The run selected the ground-truth two-compartment PK plus direct $E_{max}$ PD structure.

# *Spanish direct: A researcher with direct style*

```
Hola. Necesito que ajustes el dataset 11 del archivo data/pkpd_benchmark_data.mat. Please test the 6 PKPD
model structures: 1-compartment and 2-compartment PK, each combined with Emax, sigmoidal Emax, and
indirect response (inhibición) PD models. Use multi-start optimization (minimum 5 starts) and select the
best model using AIC.

Save results in CSV format: results/pkpd_benchmark_results.csv. Columns: dataset_id, selected_pk_model,
selected_pd_model, CL_est, Vc_est, Q_est, Vp_est, Emax_est, EC50_est, gamma_est, kin_est, kout_est,
Imax_est, IC50_est, best_WSSE, best_AIC, n_models_fitted, unique_solution, notes.
```

```
Also generate overlay plot: results/fit_dataset_11_overlay.png. Gracias.
```

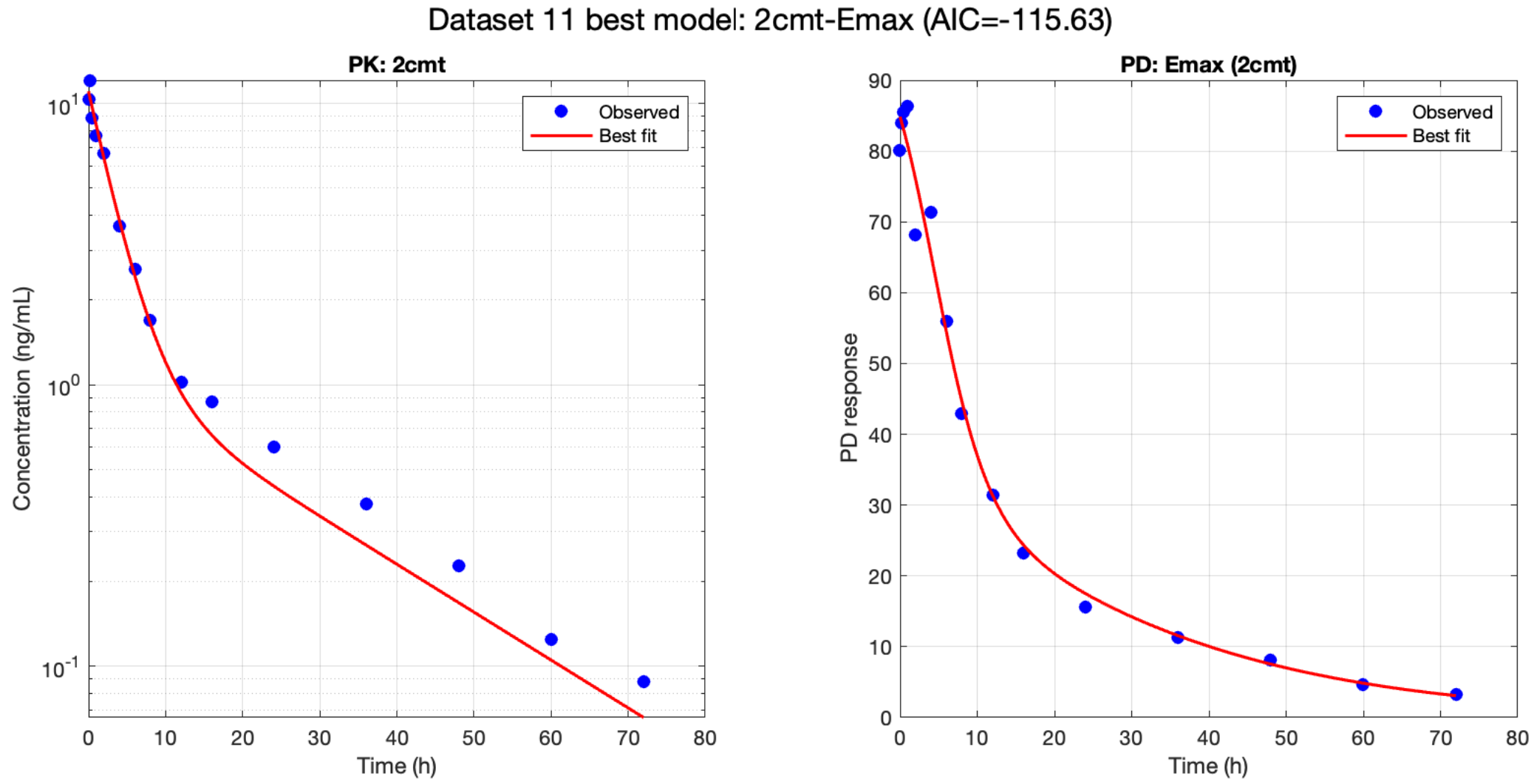

**Supplementary Figure 23.** Spanish-direct prompt result. The run selected the ground-truth two-compartment PK plus direct $E_{max}$ PD structure.

**Supplementary Table 2.** Effect of user prompt specificity on model selection

| S. No. | Prompt type | Chars | Full spec? | Exact match | Model selected |
|---|---|---|---|---|---|
| 1 | Expert concise | 514 | Yes | Yes | 2cmt + Emax |
| 2 | Expert verbose | 1383 | Yes | Yes | 2cmt + Emax* |
| 3 | Novice vague | 420 | No | Yes | 2cmt + Emax |
| 4 | Non-native English | 760 | Yes | Yes | 2cmt + Emax |
| 5 | Scandinavian formal | 774 | Yes | Yes | 2cmt + sigEmax** |
| 6 | Minimal | 134 | No | Yes | 2cmt + Emax |
| 7 | Canadian collaborative | 803 | Yes | Yes | 2cmt + Emax |
| 8 | Spanish direct | 690 | Yes | Yes | 2cmt + Emax |

*The initial expert-verbose run selected an incorrect PK structure. Repeating the identical prompt recovered the ground-truth two-compartment PK plus $E_{max}$ PD model (mean parameter error 15.0%). Independent rescoring also corrected false-negative classifications for the novice-vague and minimal variants, which selected the ground-truth model with mean parameter errors of 5.4% and 25.1%, respectively.

**The Scandinavian-formal run selected a two-compartment sigmoidal $E_{max}$ model with an estimated sigmoidicity coefficient of 1.2. Its AIC was only 1.7 lower than that of the ground-truth two-compartment $E_{max}$ model; therefore, the models were not meaningfully distinguishable under the $\Delta AIC < 2$ criterion.

# Supplementary Note 8. Literature retrieval benchmarking

To systematically evaluate the system's ability to interact with scientific literature, the Infrastructure Agent was subjected to a 17-case benchmark spanning retrieval, extraction, and multi-paper synthesis. The agent was guided by a strict system prompt instructing it to use specific API tools (e.g. `extract_pdf`, `fetch_webpage`), avoid hallucinating missing data, and cross-reference extracted metrics against the source text. The core instructions for the Infrastructure Agent are provided below, followed by the detailed benchmark queries and outcomes.

## *Supplementary Table 3: Retrieval queries*

| Case ID | Domain | Query Strategy |
|---|---|---|
| R01 | PopPK | semaglutide population pharmacokinetics |
| R02 | Bone | Riggs denosumab bone mathematical model |
| R03 | Oncology | tumor growth inhibition Simeoni pharmacokinetics |
| R04 | PMx | indirect response pharmacodynamic model Dayneka Jusko |
| R05 | PBPK | Shah Betts monoclonal antibody PBPK FcRn |
| R06 | Obesity | GLP-1 receptor agonist weight loss clinical pharmacology |
| R07 | PMx | Mager Jusko receptor mediated pharmacokinetics |
| R08 | PBPK | atenolol malnourished PBPK bioavailability |
| R09 | Obesity | SUSTAIN semaglutide subcutaneous type 2 diabetes efficacy |
| R10 | PBPK | physiologically based pharmacokinetic whole body model theophylline Khalil |

All ten open-weight runs retrieved the intended literature using the original query on the first attempt. In the closed-weight benchmark, R02, R06, and R09 required author-anchored query refinement. *The representative open-weight mean excludes R05 because its 18,558-s observed duration was dominated by a transient Azure API 403 error and repeated continuation after tool timeouts; the target paper itself was identified and relevant details downloaded within the first few minutes.

## *Supplementary Table 4: Extraction and Synthesis Queries*

| Case ID | Actual Content of Target | User Prompt | System Action* |
|---|---|---|---|
| E01 | HTTP 404 (Dead link) | Retrieve PMID 15459725 and extract: (a) the type of mathematical model, (b) the number of differential equations, (c) key PK parameters with units, (d) the study population (species, tumor type). | System reported "no records found"; did not hallucinate or fabricate content. |
| E02 | Decoy: in-vitro molecular biology text [1] | Retrieve PMID 20406950 and extract: (a) the type of mathematical model, (b) the number of state variables/ODEs, (c) key biological | System classified the target article as off-domain; declined extraction. |
| | | compartments modeled, (d) the therapeutic application. | |
| E03 | Shah and Betts 2012 [2]; mAb PBPK | Retrieve PMID 22143261 and extract: (a) the type of model, (b) the number of tissue compartments, (c) the key recycling mechanism, (d) species validated. | System accurately described the model type, tissue compartments, FcRn recycling mechanism, and validated species. |
| E04 | Decoy: Silkworm gene study [3] | Retrieve PMID 34173258 and extract: (a) the PK model structure, (b) route of administration, (c) key covariates on PK, (d) patient population. | Classified the article as an irrelevant study and declined extraction. |
| E05 | Mismatched PMID [4] | Retrieve PMID 8100098 and extract: (a) the type of PD model, (b) the number of model variants described, (c) the key equations/mechanisms, (d) example drugs used for illustration. | Recognized the erroneous PMID and declined extraction. |
| S01 | Manually review. Must cite actual Overgaard papers [5, 6]. | Search PubMed for population PK models of semaglutide (both subcutaneous and oral formulations). Compare the key PK parameters (CL/F, Vc/F, Ka, bioavailability) across the published models. Which formulation has lower bioavailability and by how much? | Autonomously retrieved three population-PK analyses, including the target papers, and correctly reasoned that oral bioavailability of approximately 0.8% versus approximately 85% for subcutaneous semaglutide—an approximately 106-fold difference—mathematically inflates apparent parameters (CL/F and $V_c$/F) despite identical systemic disposition. |
| S02 | Manual review. Must identify ≥2 real papers, produce comparison table. | Find published PBPK models for monoclonal antibodies that include FcRn recycling. Compare at least 3 papers: what tissue compartments do they include, how many ODEs, and what predicted half-life do they report for human IgG1? Summarize in a comparison table. | Autonomously identified and compared 10 PBPK models (including Shah and Betts [2], Li et al. [7], Hu et al. [8]), demonstrating calibrated confidence by explicitly distinguishing inferred ODE counts from author-reported counts. |

# References

1. Thompson, T., et al., *1,25-dihydroxyvitamin D3 enhances the apoptotic activity of MDM2 antagonist nutlin-3a in acute myeloid leukemia cells expressing wild-type p53.* Mol Cancer Ther, 2010. **9**(5): p. 1158-68.
2. Shah, D.K. and A.M. Betts, *Towards a platform PBPK model to characterize the plasma and tissue disposition of monoclonal antibodies in preclinical species and human.* J Pharmacokinet Pharmacodyn, 2012. **39**(1): p. 67-86.
3. Ye, W., et al., *Cloning and functional analysis of autophagy-related gene 7 in Bombyx mori, silkworm.* Arch Insect Biochem Physiol, 2021. **107**(4): p. e21827.
4. Motoyama, T., et al., *Differential production of interleukin 6 in human osteosarcoma cells and the possible effects on neoplastic bone metabolism.* Virchows Arch B Cell Pathol Incl Mol Pathol, 1993. **63**(5): p. 277-81.
5. Overgaard, R.V., et al., *Population Pharmacokinetics of Semaglutide for Type 2 Diabetes.* Diabetes Ther, 2019. **10**(2): p. 649-662.
6. Overgaard, R.V., et al., *Clinical Pharmacokinetics of Oral Semaglutide: Analyses of Data from Clinical Pharmacology Trials.* Clin Pharmacokinet, 2021. **60**(10): p. 1335-1348.
7. Li, L., et al., *Simulation of monoclonal antibody pharmacokinetics in humans using a minimal physiologically based model.* AAPS J, 2014. **16**(5): p. 1097-109.
8. Hu, S., A. Datta-Mannan, and D.Z. D'Argenio, *Physiologically Based Modeling to Predict Monoclonal Antibody Pharmacokinetics in Humans from in vitro Physiochemical Properties.* MAbs, 2022. **14**(1): p. 2056944.
9. Kir, F., S. Sahin, and W.J. Jusko, *Minimal Physiologically-Based Pharmacokinetic Modeling of Atenolol and Metoprolol Absorption in Malnourished Rats.* Eur J Drug Metab Pharmacokinet, 2025. **50**(3): p. 251-263.